\documentclass[preprint,pteplogo]{ptephy_v2}

\preprintnumber{2410-14664} 

\usepackage{hyperref}

\usepackage[T1]{fontenc} 
\usepackage{tikz-feynhand}
\usepackage{graphicx}
\usepackage{xcolor}
\usepackage{caption,subcaption}
\usepackage{mathrsfs,mathtools}
\usepackage{physics,amssymb}
\usepackage{bm}
\usepackage{braket}
\usepackage{listings}
\usepackage{cases}
\usepackage{comment}
\usepackage{soul}
\usepackage{cancel}
\usepackage{cases}
\usepackage[utf8]{inputenc}
\usepackage{url}
\usepackage[normalem]{ulem}
\usepackage{xspace}
\usepackage{siunitx}

\hypersetup{colorlinks=true
,urlcolor=DARKBLUE
,anchorcolor=DARKBLUE
,citecolor=DARKBLUE
,filecolor=DARKBLUE
,linkcolor=DARKBLUE
,menucolor=DARKBLUE
,linktocpage=true
,pdfproducer=medialab
,pdfa=true
}

\usepackage{xcolor}
\definecolor{bleudefrance}{rgb}{0.19, 0.55, 0.91}
\definecolor{desyblue}{HTML}{009EE2}
\definecolor{desyorange}{HTML}{FD8800}
\definecolor{dark_red}{rgb}{0.7, 0., 0.}
\definecolor{light_pink}{rgb}{1,0.4,0.4}
\definecolor{lblue}{rgb}{0.384602,0.117763,0.973947}
\usepackage{color}
\input{colordvi.tex}
\allowdisplaybreaks[1]
\usepackage{url}
\usepackage{hyperref}
\hypersetup{
colorlinks=false,
hidelinks}
\newcommand*{\D}{{\rm d}}
\newcommand*{\mpl}{M_{\rm Pl}}

\definecolor{MONZA}{HTML}{CF000F}
\definecolor{DARKBLUE}{HTML}{00008b}
\definecolor{DARKMAGENTA}{HTML}{8b008b}
\definecolor{DARKCYAN}{HTML}{00cfc0}

{}

\begin{document}
\preprintnumber{RUP-24-20}
\title{Stochastic gravitational wave background anisotropies from inflation with non-Bunch-Davies states}


\author[a]{Shingo Akama}
\author[b,c]{Shin'ichi Hirano}
\author[d,e]{Shuichiro Yokoyama}

\affil[a]{Faculty of Physics, Astronomy and Applied Computer Science, Jagiellonian University, 30-348 Krakow, Poland \email{shingo.akama@uj.edu.pl}}
\affil[b]{National Institute of Technology, Oyama College
771 Nakakuki, Oyama-shi, Tochigi 323-0806, Japan}
\affil[c]{Department of Physics, Rikkyo University, Toshima, Tokyo 171-8501, Japan \email{hirano.s.aile@oyama-ct.ac.jp}}
\affil[d]{Kobayashi Maskawa Institute, Nagoya University, 
Chikusa, Aichi 464-8602, Japan}
\affil[e]{Kavli Institute for the Physics and Mathematics of the Universe (WPI), UTIAS, The University of Tokyo, Kashiwa, Chiba 277-8583, Japan \email{shu@kmi.nagoya-u.ac.jp}}




\begin{abstract}
It is known that stochastic gravitational wave backgrounds (SGWBs) have anisotropies generated by squeezed-type tensor non-Gaussianities originating from scalar-tensor-tensor (STT) and tensor-tensor-tensor cubic interactions. While the squeezed tensor non-Gaussianities in the standard slow-roll inflation with the Bunch-Davies vacuum state are suppressed due to the so-called consistency relation, those in extended models with the violation of the consistency relation can be enhanced. Among such extended models, we consider the inflation model with the non-Bunch-Davies state that is known to enhance the squeezed tensor non-Gaussianities. We explicitly formulate the primordial STT bispectrum induced during inflation in the context of Horndeski theory with the non-Bunch-Davies state and show that the induced SGWB anisotropies can be enhanced. We then discuss the detectability of those anisotropies in future gravitational wave experiments.
\end{abstract}

\subjectindex{E02,E03,E81}

\maketitle

\section{Introduction}

Recently, 
several international collaborations for the Pulsar Timing Array (PTA) gravitational wave (GW) experiments such as North American Nanohertz Observatory for Gravitational Waves (NANOGrav)~\cite{McLaughlin:2013ira, Brazier:2019mmu}
have reported evidence of nHz stochastic gravitational wave background (SGWB)~\cite{NANOGrav:2023gor, EPTA:2023fyk, Reardon:2023gzh, Xu:2023wog}.
Furthermore, future ground- and space-based missions at higher frequency bands, e.g. Cosmic Explorer (CE)~\cite{Reitze:2019iox}, Einstein Telescope (ET)~\cite{Punturo:2010zz}, Laser Interferometer Space Antenna (LISA)~\cite{LISA:2017pwj, LISACosmologyWorkingGroup:2022jok}, Taiji~\cite{Hu:2017mde}, the Taiji-LISA network~\cite{Ruan:2020smc, Wang:2021uih}, Decihertz Gravitational Wave Observatory (DECIGO)~\cite{Kawamura:2006up, Kawamura:2020pcg}, the Big Bang Observer (BBO)~\cite{Crowder:2005nr}, etc., are expected to report other evidence for the SGWB at different frequency ranges. 
Although the most plausible source of the SGWB is the superposition of the GWs from astrophysical compact objects such as black holes (see, e.g. Ref.~\cite{NANOGrav:2023hfp}), various cosmological origins of the SGWB, in particular, from the early universe with new physics have also been widely discussed (see, e.g. Ref.~\cite{NANOGrav:2023hvm}).
It is, therefore, worth discussing the possibilities of specifying the source of SGWB and testing new physics in the early universe
with the PTA experiments and the ground- and space-based missions. 

As one of the things that can help identify the sources, the anisotropies of SGWB have been recently investigated~\cite{Contaldi:2016koz, Cusin:2017fwz, Ricciardone:2017kre,Dimastrogiovanni:2018uqy,Dimastrogiovanni:2019bfl,Bartolo:2019oiq, Bartolo:2019zvb,Bartolo:2019yeu,Adshead:2020bji,Malhotra:2020ket,Dimastrogiovanni:2021mfs,Dimastrogiovanni:2022afr,Dimastrogiovanni:2022eir,Orlando:2022rih,Malhotra:2022ply,Li:2023qua, Li:2023xtl, Yu:2023jrs}. In addition to the anisotropies associated with the metric fluctuations in the late time universe (see, e.g. Refs.~\cite{Contaldi:2016koz, Cusin:2017fwz, Bartolo:2019oiq, Bartolo:2019yeu}) known as the so-called Sachs-Wolfe (SW) and integral SW effects for the cosmic microwave background (CMB) anisotropies,\footnote{See also Refs.~\cite{Bartolo:2019zvb, Dimastrogiovanni:2022eir, Malhotra:2022ply, Li:2023qua, Li:2023xtl, Yu:2023jrs} where the anisotropies of secondary GWs sourced by the late-time perturbation have been studied.} the SGWB anisotropies can be sourced by the primordial scalar-tensor-tensor (STT) and tensor-tensor-tensor (TTT) cubic interactions in the squeezed configuration~\cite{Ricciardone:2017kre, Dimastrogiovanni:2018uqy, Dimastrogiovanni:2019bfl, Adshead:2020bji, Malhotra:2020ket, Dimastrogiovanni:2021mfs, Dimastrogiovanni:2022afr, Orlando:2022rih}.
In the present paper, we discuss the SGWB anisotropies 
induced by the primordial STT correlations generated during inflation. 
In the standard single scalar inflation, the consistency relation prohibits any enhancements of the primordial non-Gaussianities in the squeezed configuration~\cite{Maldacena:2002vr}.
On the other hand, in extended models with its violation, those non-Gaussianities can be enhanced, and then the SGWB anisotropies in such non-standard inflationary models have been studied in Refs.~\cite{Ricciardone:2017kre, Dimastrogiovanni:2018uqy, Dimastrogiovanni:2019bfl, Adshead:2020bji, Malhotra:2020ket, Dimastrogiovanni:2021mfs, Dimastrogiovanni:2022afr, Orlando:2022rih}. Here, we consider inflation with non-Bunch-Davies states among the extended models wherein the consistency relation is violated.

Inflation with the non-Bunch-Davies states has been shown to potentially enhance the primordial non-Gaussianities as a consequence of the violation of the consistency relation~\cite{Chen:2006nt, Holman:2007na, Xue:2008mk, Meerburg:2009ys, Meerburg:2009fi, Chen:2010xka, Meerburg:2010ks, Chen:2010bka, Meerburg:2010rp, Agullo:2010ws, Ashoorioon:2010xg, Ganc:2011dy, LopezNacir:2011kk, Ganc:2012ae, Agarwal:2012mq, Gong:2013yvl, Flauger:2013hra, Aravind:2013lra, Ashoorioon:2013eia, Brahma:2013rua, Bahrami:2013isa, Kundu:2013gha, Emami:2014tpa, Zeynizadeh:2015eia, Meerburg:2015yka, Ashoorioon:2016lrg, Shukla:2016bnu, Ashoorioon:2018sqb, Brahma:2019unn, Akama:2020jko, Naskar:2020vkd, Ragavendra:2020vud, Kanno:2022mkx, Ghosh:2022cny, Gong:2023kpe, Ghosh:2023agt, Akama:2023jsb, Cielo:2023enz,Peng:2024eok, Ansari:2024pgq, Christodoulidis:2024ric}. In particular, the bispectra of the scalar and tensor perturbations are enhanced in proportion to powers of $k_S/k_L$ where $k_S$ and $k_L$ respectively denote the comoving wavenumber of the short- and long-wavelength modes in the squeezed limit, i.e. $k_L/k_S \to 0$,
and then it is naively expected that such a $k_S/k_L$ enhancement could efficiently amplify the SGWB anisotropies.
However, since this characteristic enhancement of the primordial non-Gaussianities due to the non-Bunch-Davies states is basically induced by the cubic interactions on subhorizon scales,
the short-wavelength modes may suffer from a so-called Trans-Planckian problem~\cite{Martin:2000xs} and a strong coupling problem in the squeezed limit $k_S \gg k_L$
when $k_S$ and $k_L$ are respectively taken to be the target frequency of observed SGWBs and the scale of interest for their anisotropies on large scales.
To obtain proper predictions for the SGWB anisotropies from the inflation with non-Bunch-Davies states, we first derive a condition to avoid the Trans-Planckian and strong coupling problems and then obtain the primordial STT bispectrum from the inflation with non-Bunch-Davies states under that condition. Also, for testing the SGWB anisotropies from inflation, it is convenient to study those in a general framework of inflation, and hence we work with the Horndeski theory, which is the most general scalar-tensor theory whose equations of motion are second order~\cite{Horndeski:1974wa, Deffayet:2011gz, Kobayashi:2011nu} (see also Ref.~\cite{Kobayashi:2019hrl} for a review).
As a result, we will show that the resultant non-Gaussianities and SGWB anisotropies from the general single-field inflation can be enhanced in a different way than the $k_S/k_L$ enhancement.
We also discuss whether the SGWB anisotropies can be detected with actual gravitational wave experiments. The DECIGO/BBO experiments are designed to test inflation models predicting an almost scale-invariant tensor spectrum with the tensor-to-scalar ratio being $\mathcal{O}(10^{-2})$, whereas the recent NANOGrav data can be explained from inflation with a certain blue-tilted tensor spectrum. The power spectrum of the non-Bunch-Davies tensor modes generally has scale dependence via a Bogoliubov coefficient, and hence it is possible for inflation with the non-Bunch-Davies states to explain the recent results of nHz SGWB~\cite{Choudhury:2023kam}. 
In light of these GW observations, we consider two configurations of the Bogoliubov coefficient. One is a scale-independent coefficient. This configuration leads to an almost scale-invariant tensor spectrum; the SGWB originating from which can potentially be tested by DECIGO/BBO. The other is a coefficient that peaks at a certain scale $k=k_*$. The resultant tensor power spectrum has a peak at $k=k_*$ as well, and hence, properly choosing the value of $k_*$, the resultant GW can explain the recent PTA data. 
We would also like to investigate the cross-correlation between the CMB and the SGWB anisotropies~\cite{Adshead:2020bji}.

This paper is organized as follows. In the following section, we briefly review our setup and primordial power spectra of scalar and tensor perturbations with the non-Bunch-Davies states. In Sec.~\ref{Sec: cross-bispectrum}, we first explain the Trans-Planckian and strong coupling problems for the short-wavelength GWs and clarify the condition to avoid these problems. Then, we compute the primordial STT bispectrum with the non-Bunch-Davies states under that condition. In Sec.~\ref{Sec: Bogoliubov-coeff}, we introduce two phenomenological models for the Bogoliubov coefficient such that the tensor power spectrum is nearly scale-invariant or has a peak at a certain scale. In Sec.~\ref{Sec: SGWB}, after reviewing the impacts of the primordial non-Gaussianities on the SGWB anisotropies, we evaluate the auto-correlation function of those and the cross-correlation one between those and CMB anisotropies. Then, we discuss the detectability of those correlation functions. The summary of this paper is drawn in Sec.~\ref{Sec: conclusion}.

\section{Primordial power spectra with non-Bunch-Davies states}\label{Sec: power-spectrum}

To explore the impact of primordial non-Gaussianities induced by the non-Bunch-Davies states on the SGWB,
it is useful to formulate them in a general framework of single-field inflation.
We work with the Horndeski theory~\cite{Horndeski:1974wa, 
 Deffayet:2011gz, Kobayashi:2011nu}, which is the most general scalar-tensor theory whose equations of motion are second order. The action takes the following form~\cite{Kobayashi:2011nu}:
\begin{align}
S&=\int\D^4 x\sqrt{-g}\biggl[G_2(\phi,X)-G_3(\phi,X)\Box\phi+G_4(\phi,X)R+G_{4X}[(\Box\phi)^2-(\nabla_{\mu}\nabla_{\nu}\phi)^2]\notag\\
&\quad +G_5(\phi,X)G_{\mu\nu}\nabla^{\mu}\nabla^{\nu}\phi-\frac{1}{6}G_{5X}[(\Box\phi)^3-3\Box\phi(\nabla_{\mu}\nabla_{\nu}\phi)^2+2(\nabla_{\mu}\nabla_{\nu}\phi)^3]\biggr], \label{eq: Horndeski action}
\end{align}
with $G_i(\phi,X)$ being arbitrary functions of the scalar field $\phi$ and its canonical kinetic term $X:=-g^{\mu\nu}\partial_\mu\phi\partial_\nu\phi/2$. 
Imposing appropriate conditions on the arbitrary functions, the scalar field drives the slow-roll inflation as an inflaton~\cite{Kobayashi:2011nu}.

The perturbed metric around a spatially flat Friedmann-Lema\^{i}tre-Robertson-Walker spacetime under the unitary gauge in which the spatial fluctuation of the scalar field vanishes, i.e. $\phi = \phi(t)$, is of the form
\begin{align}
\D s^2=-N^2\D t^2+g_{ij}(\D x^i+N^i\D t)(\D x^j+N^j\D t), \label{eq: perturbed-metric}
\end{align}
where
\begin{align}
N&=1+\alpha,\ N_i=\partial_i\beta,\\
g_{ij}&=a^2 e^{2\zeta}\biggl(\delta_{ij}+h_{ij}+\frac{1}{2}h_{ik}h_{kj}+\cdots\biggr), \label{eq: def-of-perturbations}
\end{align}
with $a(t)$ being a scale factor. Here, $\alpha,~\beta,$ and $\zeta$ are scalar perturbations, and $h_{ij}$ means transverse-traceless tensor perturbations.
We hereafter use a dot as a derivative with respect to $t$, and the Hubble parameter is defined by $H:=\dot a/a$.

\subsection{Quadratic action and primordial power spectra with non-Bunch-Davies states}

Expanding the action Eq.~(\ref{eq: Horndeski action}) up to the quadratic order in the perturbations and substituting the solutions of $\alpha$ and $\beta$ derived from the constraint equations into the perturbed action, one obtains~\cite{Kobayashi:2011nu}:
\begin{align}
S^{(2)}_\zeta&=\int\D t\D^3x a^3\biggl[\mathcal{G}_S\dot\zeta^2-\frac{\mathcal{F}_S}{a^2}(\partial_i\zeta)^2\biggr], \label{eq: quad-scalar}\\
S^{(2)}_h&=\frac{1}{8}\int\D t\D^3x a^3\biggl[\mathcal{G}_T\dot h_{ij}^2-\frac{\mathcal{F}_T}{a^2}(\partial_k h_{ij})^2\biggr], \label{eq: quad-tensor}
\end{align}
where $\mathcal{G}_S,~\mathcal{F}_S,~\mathcal{G}_T$, and $\mathcal{F}_T$ are given in terms of $\phi$, $X$, the arbitrary functions $G_i(\phi,X)$, and their derivatives with respect to $\phi$ and $X$, and the explicit expressions of these functions are given in Appendix~\ref{App: perturbed-action}. Throughout this paper, we assume a slow-roll inflation with $\mathcal{G}_S,~\mathcal{F}_S,~\mathcal{G}_T,~\mathcal{F}_T\simeq {\rm const.}$ in a quasi-de Sitter spacetime ($H\simeq {\rm const.}$). Specifically, we introduce the slow-roll parameters as
\begin{align}
\epsilon:=-\frac{\dot H}{H^2},\ g_S:=\frac{\dot{\mathcal{G}}_S}{H\mathcal{G}_S},\ f_S:=\frac{\dot{\mathcal{F}}_S}{H\mathcal{F}_S},\ g_T:=\frac{\dot{\mathcal{G}}_T}{H\mathcal{G}_T},\ f_T:=\frac{\dot{\mathcal{F}}_T}{H\mathcal{F}_T},
\end{align}
and assume $\epsilon,~g_S,~f_S,~g_T,~f_T\ll1$ as slow-roll approximations. 

We then quantize the scalar and tensor perturbations. The Fourier transformations of those are defined by
\begin{align}
\zeta(t,{\bf x})&=\int\frac{\D^3k}{(2\pi)^3}\tilde\zeta(t,{\bf k})e^{i\bf k\cdot\bf x},\\
h_{ij}(t,{\bf x})&=\int\frac{\D^3k}{(2\pi)^3}\tilde h_{ij}(t,{\bf k})e^{i\bf k\cdot\bf x}.
\end{align}
In Fourier space, both perturbations are quantized as
\begin{align}
\tilde{\zeta} \to~ \hat\zeta(t,{\bf k})&=u_k(t) \hat a_{\bf k}+u_k^*(t)\hat a^\dagger_{-{\bf k}},\\
\tilde{h}_{ij} \to~ \hat h_{ij}(t,{\bf k})&=\sum_s \biggl[v_k(t)e^{(s)}_{ij}({\bf k})\hat a^{(s)}_{\bf k}+v^*_k(t)e^{(s)*}_{ij}(-{\bf k})\hat a^{(s)\dagger}_{-{\bf k}}\biggr],
\end{align}
where  $e^{(s)}_{ij}(\bf k)$ denotes the polarization tensor with the helicity states $s=\pm 2$, satisfying the transverse-traceless conditions, i.e. $k_i e^{(s)}_{ij}({\bf k})=\delta_{ij}e^{(s)}_{ij}({\bf k})=0$.
The creation and annihilation operators satisfy the standard commutation relations,
\begin{align}
[\hat a_{\bf k},\hat a^\dagger_{{\bf k}'}]&=(2\pi)^3\delta({\bf k}-{\bf k}'),\\
[\hat a^{(s)}_{\bf k},\hat a^{(s')\dagger}_{{\bf k}'}]&=(2\pi)^3\delta_{ss'}\delta({\bf k}-{\bf k}'),\\
{\rm others}&=0.
\end{align}
The equations of motion for the perturbations in real space are of the form,
\begin{align}
E^s&:=\partial_t(a^3\mathcal{G}_S\dot\zeta)-a\mathcal{F}_S\partial^2\zeta=0,\\
E^h_{ij}&:=\partial_t(a^3\mathcal{G}_T\dot h_{ij})-a\mathcal{F}_T\partial^2 h_{ij}=0.
\end{align}
On the quasi-de Sitter background and under the slow-roll approximations, at the leading order, general solutions of the equations of motion for the mode functions 
can be obtained as
\begin{align}
u_k&=\alpha_ku_{k,({\rm BD})}+\beta_ku^*_{k,({\rm BD})},\\
v^{(s)}_k&=\alpha^{(s)}_kv_{k,({\rm BD})}+\beta^{(s)}_kv^*_{k,({\rm BD})},
\end{align}
where $\alpha_k,~\beta_k,~\alpha^{(s)}_k$, and $\beta^{(s)}_k$ are Bogoliubov coefficients, and $u_{k,({\rm BD})}$ and $v_{k,({\rm BD})}$ denote the mode functions with the Bunch-Davies state which are respectively given as~\cite{Kobayashi:2011nu}:
\begin{align}
u_{k,({\rm BD})}&=\frac{iH}{2\sqrt{\mathcal{F}_Sc_sk^3}}(1+ic_sk\eta)e^{-ic_sk\eta},\\
v_{k,({\rm BD})}&=\frac{i\sqrt{2}H}{\sqrt{\mathcal{F}_Tc_hk^3}}(1+ic_hk\eta)e^{-ic_hk\eta},
\end{align}
where we introduced the conformal time $\eta:=\int\D t/a$ and the propagation speed of each mode $c_s^2=\mathcal{F}_S/\mathcal{G}_S,~c_h^2=\mathcal{F}_T/\mathcal{G}_T$.
The Bogoliubov coefficients satisfy the following normalization conditions:
\begin{align}
|\alpha_k|^2-|\beta_k|^2&=1,\\
|\alpha^{(s)}_k|^2-|\beta^{(s)}_k|^2&=1,
\label{eq: normalization-bogo}
\end{align}
which are derived from the Wronskian conditions. 
The dimensionless primordial power spectra of the scalar and tensor perturbations are respectively defined by
\begin{align}
\langle\hat\zeta({\bf k})\hat\zeta({\bf k}')\rangle&=(2\pi)^3\delta({\bf k}+{\bf k}')\frac{2\pi^2}{k^3}\mathcal{P}_\zeta,\\
\langle\hat\xi^{(s)}({\bf k})\hat\xi^{(s')}({\bf k}')\rangle&=(2\pi)^3\delta_{ss'}\delta({\bf k}+{\bf k}')\frac{\pi^2}{k^3}\mathcal{P}^{(s)}_h,
\end{align}
where $\hat\xi^{(s)}({\bf k}):=\hat h_{ij}({\bf k})e^{(s)*}_{ij}({\bf k})$.
Then, the power spectra on superhorizon scales ($-c_sk\eta\ll1$ and $-c_h k\eta \ll 1$), can be obtained from the mode functions as
\begin{align}
\mathcal{P}_\zeta&=\frac{1}{8\pi^2}\frac{H^2}{\mathcal{F}_Sc_s}|\alpha_k-\beta_k|^2, \label{eq: power-scalar}\\
\mathcal{P}^{(s)}_h&=\frac{1}{\pi^2}\frac{H^2}{\mathcal{F}_Tc_h}|\alpha^{(s)}_k-\beta^{(s)}_k|^2, 
\label{eq: power-tensor}
\end{align}
Then, up to the leading order in the slow-roll parameters, the spectral indices of the scalar and tensor power spectra are derived as
\begin{align}
n_s-1&:=\frac{\D\ln \mathcal{P}_\zeta}{\D\ln k}\simeq-2\epsilon+\frac{g_S}{2}-\frac{3}{2}f_S+\frac{\D}{\D \ln k}\ln|\alpha_k-\beta_k|^2, \label{eq: tilt-scalar}\\
n_t&:=\frac{\D\ln \mathcal{P}_h}{\D\ln k}\simeq-2\epsilon+\frac{g_T}{2}-\frac{3}{2}f_T+\frac{\D}{\D \ln k}\ln\sum_s|\alpha^{(s)}_k-\beta^{(s)}_k|^2, \label{eq: tilt-tensor}
\end{align}
where $\mathcal{P}_h = \sum_s \mathcal{P}^{(s)}_h$.
By taking the limit, $\alpha_k=1,~\beta_k=0,~\alpha^{(s)}_k=1$, and $\beta^{(s)}_k=0$, Eqs.~(\ref{eq: power-scalar})--(\ref{eq: tilt-tensor}) reproduce the results in the slow-roll inflation with the Bunch-Davies state~\cite{Kobayashi:2011nu}. In the main text, we hereafter set $\alpha_k=1$ and $\beta_k=0$ (i.e. assume the Bunch-Davies scalar modes) for simplicity, and we solely study the impact of the non-Bunch-Davies tensor modes on the SGWB and its anisotropies. (See Ref.~\cite{Fumagalli:2021mpc} for the impacts of primordial GWs induced by non-Bunch-Davies scalar modes on the SGWB.)

\subsection{On the scale-dependence of the tensor power spectrum with non-Bunch-Davies state}
\label{Sec: Bogoliubov-coeff}

The last term in Eq.~(\ref{eq: tilt-tensor}) is peculiar to the excited tensor modes, and $k$-dependence of the Bogoliubov coefficients can yield a variety of non-standard predictions, e.g. a highly blue-tilted spectrum~\cite{Ashoorioon:2014nta,Choudhury:2023kam}.
Such a blue-tilted spectrum can potentially be tested by PTA experiments such as NANOGrav, as has been studied in Ref.~\cite{Choudhury:2023kam} where a power-law $k$-dependence of the power spectrum was assumed.
On the other hand, the nearly scale-invariant spectrum also has a chance to be tested by planned space interferometers such as DECIGO/BBO.
Thus, here, 
we introduce phenomenological models for the Bogoliubov coefficients to realize sufficient amplitude of the SGWB for detection in both frequency bands. (See, e.g. Refs.~\cite{Danielsson:2002kx,Goldstein:2002fc,Alberghi:2003am,Broy:2016zik,Cielo:2022vmo} where impacts of non-Bunch-Davies states on scalar/tensor power spectra have been studied with different Bogoliubov coefficients.)

\subsubsection{Peaked spectrum}
\label{sec:intro peaked}

To explore the possible enhancement at a certain scale, we consider an example in which the primordial tensor power spectrum has a peak at a scale $k=k_*$. Let us first impose the following ansatz on the Bogoliubov coefficients:
\begin{align}
\alpha^{(s)}_k&=1+i{\rm Im}[\alpha^{(s)}_k],\\
\beta^{(s)}_k&={\rm Re}[\beta^{(s)}_k].
\end{align}
From the normalization condition Eq.~(\ref{eq: normalization-bogo}), one can derive 
\begin{align}
({\rm Im}[\alpha^{(s)}_k])^2=({\rm Re}[\beta^{(s)}_k])^2. \label{eq: bogo-alpha}
\end{align}
We hereafter assume ${\rm Im}[\alpha^{(s)}_k]={\rm Re}[\beta^{(s)}_k]$ for simplicity.  Then, to realize the peak, we further consider
\begin{align}
{\rm Re}[\beta^{(s)}_k]=A\times \exp\biggl[-\frac{(\ln{k/k_*})^2}{\Delta^2}\biggr], \label{eq: bogo-beta}
\end{align}
where $A$ and $\Delta$ are dimensionless parameters that control how large and how sharp the peak is. Here, the above ansatz covers various types of the GW spectrum such as an almost flat spectrum and a weaker or broader peaked one. However, among those, one needs to take $A$ to be much smaller than unity to get the flat spectrum, which suppresses the enhancement of anisotropies as will be shown later. Therefore, as an exception, we consider a different ansatz leading to the flat spectrum that can enhance the anisotropies in the following subsection. 

For $A=-3000$ and $\Delta=0.45$, we can realize a blue-tilted tensor spectrum
with 
$n_t\simeq 2$ 
for $k \lesssim k_*$,
and an $\mathcal{O}(10^{7})$ enhancement in the power spectrum at a reference scale $k=k_*$.
Ref.~\cite{Choudhury:2023kam} mentioned that 
the nHz SGWB detected by recent PTA experiments~\cite{NANOGrav:2020bcs,Goncharov:2021oub,EPTA:2021crs,EPTA:2023fyk,EPTA:2023sfo,EPTA:2023akd,EPTA:2023gyr,EPTA:2023xxk,EuropeanPulsarTimingArray:2023egv,Antoniadis:2022pcn}
can be explained by the single-field inflation with the non-Bunch-Davies vacuum state if choosing $k_* \sim 10^8 {\rm Mpc}^{-1}$. Here, one might consider different values of $\Delta$ to realize a different sharpness of the GW spectrum with enhanced anisotropies. Note that, as we will see later, we have to be careful about the backreaction from the excited states on the inflationary background dynamics (see, e.g. Ref.~\cite{Akama:2020jko}). For $\Delta \gtrsim 1$ to realize a wider peak, more modes contribute to the backreaction, and thus we will get a more stringent constraint on the non-Bunch-Davies states compared to the above choice. 
In fact, 
as will be shown later, one cannot arbitrarily increase the value of $\eta_0$, which is introduced as an excitation time of the Bunch-Davies tensor modes, due to Trans-Planckian and strong coupling issues, and the maximum possible value of $-k\eta_0$ for $\Delta=0.45$ is almost the same as that determined by requiring the absence of those issues. Therefore, to discuss the highest detectability of anisotropies by the PTA experiments, we do not consider the other values.

\subsubsection{Nearly scale-invariant spectrum}
\label{sec:intro scaleinv}
As we mentioned before, DECIGO can potentially test a nearly scale-invariant spectrum. Thus, as the minimum setup for the case of DECIGO, we consider a nearly flat spectrum by employing the following ansatz:\footnote{Here, we assumed a pure imaginary $\beta^{(s)}_k$. This is because the scale-invariant case with a pure real $\beta^{(s)}_k$ yields a leading-order contribution of $\mathcal{O}(B^2)$ in Eqs.~(\ref{eq: block-bogoliubov-1}) and~(\ref{eq: block-bogoliubov-2}). As will be explained later, one has $B\ll1$ due to the backreaction constraint. Therefore, 
we need a pure imaginary $\beta^{(s)}_k$ that leads to the leading-order contribution of $\mathcal{O}(B)$ in Eqs.~(\ref{eq: block-bogoliubov-1}) and~(\ref{eq: block-bogoliubov-2}).}
\begin{align}
\alpha^{(s)}_k&=1+iB,\\
\beta^{(s)}_k&=iB,
\end{align}
where $B$ is real and constant. In this case, 
the last term in Eq.~(\ref{eq: tilt-tensor}) is zero, and
the spectral index of the tensor power spectrum is given by the slow-roll parameters.

\section{Non-Bunch-Davies bispectrum}\label{Sec: cross-bispectrum}

As we will see in section~\ref{Sec: SGWB}, the SGWB anisotropies on large scales are sensitive to the primordial non-Gaussianity which gives the coupling between short- and long-wavelength modes (see, e.g. Refs.~\cite{Dimastrogiovanni:2021mfs, Orlando:2022rih}). At the leading order, such a coupling can be characterized by the primordial bispectrum in the squeezed limit. 
Although the tensor auto-bispectrum in the squeezed limit can be a source of the anisotropies of primordial GWB, due to the tight observational constraint on the primordial tensor modes on CMB scales the contribution from the tensor auto-bispectrum might be smaller than that from the STT cross-bispectrum. 
Then, here, we focus on the primordial STT bispectrum in the squeezed limit, i.e. the coupling between long-wavelength scalar modes and short-wavelength tensor modes.

\subsection{Trans-Planckian and strong coupling problems}

Before formulating the STT bispectrum, we
mention the Trans-Planckian issue~\cite{Martin:2000xs} and a strong coupling issue in the inflation models with the non-Bunch-Davies state. 


In general, in the inflationary era, the physical momenta of each mode could exceed the Planck scale.
In particular, this issue may become more serious in the situation considered here, where the large hierarchy between the long- and short-wavelength modes exists.
In addition, Ref.~\cite{Kunimitsu:2015faa} pointed out that in the Horndeski theory the strong coupling scale should generally appear below the Planck scale and 
it is estimated as
\begin{align}
\Lambda \simeq (\sqrt{\epsilon}\mpl H^2)^{1/3} \ll \mpl,
\label{eq: cutoff}
\end{align}
where $\mathcal{G}_S/\mpl^2, \mathcal{F}_S/\mpl^2 \sim \epsilon \ll 1$ and $H/\mpl \ll 1$ are assumed. It should be noted that, depending on the choice of the arbitrary functions in Horndeski's action~\eqref{eq: Horndeski action}, a higher cutoff scale can be obtained. Hereafter, we employ the lowest strong coupling scale given by Eq.~\eqref{eq: cutoff}, which gives the most conservative condition on the excitation time of the tensor modes, $\eta_0$.

Thus, we need to require that the physical momenta, in particular, of the short-wavelength modes do not exceed these scales.
We discuss this issue separately in two situations.
First, let us consider the case where both short- and long-wavelength modes are on subhorizon scales at a certain time
during inflation.
The condition that the long-wavelength modes are on subhorizon scales at $\eta=\eta_0$ is expressed as
\begin{align}
1 < -k_L\eta_0 \,\left(\,\ll - k_S \eta_0\,\right), \label{eq:ineq-TransPlanckian}
\end{align}
where $k_S$ and $k_L$ denote the comoving wavenumber of the short- and 
long-wavelength modes, respectively. 
In the quasi-de Sitter background (i.e. $a\simeq -(H\eta)^{-1}$), Eq.~(\ref{eq:ineq-TransPlanckian}) turns into
\begin{align}
k_{{\rm phys},L} > H_{\rm inf},~ k_{{\rm phys},S} > H_{\rm inf} \times \frac{k_S}{k_L},~
\end{align}
at $\eta = \eta_0$. As for the strong coupling issue, it should be required that
\begin{align}
1 > \frac{k_{{\rm phys},S}}{\Lambda}~,    
\end{align}
and hence we can obtain the requirement for the hierarchy between the short- and long-wavelength modes as
\begin{align}
 \frac{\Lambda}{H_{\rm inf}}  >  \frac{k_S}{k_L}~.  
\end{align}
By using the expression for the cutoff scale in generalized G-inflation \eqref{eq: cutoff}, we have
\begin{align}
 \left(\frac{\sqrt{\epsilon} \mpl}{H_{\rm inf}}\right)^{1/3}  >  \frac{k_S}{k_L}~, 
\end{align}
and this inequality can be estimated as
\begin{align}
O(10) \times \left(\frac{\mathcal{P}_\zeta}{2 \times 10^{-9}}\right)^{-1/6} > \frac{k_S}{k_{\rm CMB}}~,   
\end{align}
where we have used the expression for the power spectrum of the curvature perturbations \eqref{eq: power-scalar}
and assumed $\mathcal{G}_S/\mpl^2, \mathcal{F}_S/\mpl^2 \sim \epsilon$ and $|\alpha_{k_{\rm CMB}}-\beta_{k_{\rm CMB}}|^2\sim 1$. Here, we take $k_L$ as the CMB scale.
This inequality means that the detection of the inflationary SGWB
would suffer from the strong coupling problem of the short-wavelength modes at $\eta = \eta_0$
for $k_S > O(10) \times k_{\rm CMB}$.
To avoid this problem, let us consider the case that relaxes the starting condition of the above argument based on
Eq.~\eqref{eq:ineq-TransPlanckian}, in such a way that the tensor modes with $k_S$ got excited from the Bunch-Davies state at $\eta=\eta_0$ when the long-wavelength modes of $k_L$ (typically, CMB scale) were already on superhorizon scales, while the short-wavelength modes were still on subhorizon scales:
\begin{align}
-k_S\eta_0>1\gg-k_L\eta_0. \label{eq:new-ineq-TransPlanckian}
\end{align}
The strong coupling problem requires
\begin{align}
\frac{\Lambda}{H_{\rm inf}} > -k_S\eta_0,   \label{eq: constraint-keta-cutoff}
\end{align}
at $\eta = \eta_0$,
where we have used $a\simeq - (H\eta)^{-1}$.
By employing Eq.~\eqref{eq: cutoff}, we have
\begin{align}
 \left(\frac{\sqrt{\epsilon} \mpl}{H_{\rm inf}}\right)^{1/3}  >   -k_S\eta_0~, 
\end{align}
and this inequality can be estimated as
\begin{align}
O(10) \times \left(\frac{\mathcal{P}_\zeta}{2 \times 10^{-9}}\right)^{-1/6} >  -k_S\eta_0.   \label{eq: constraint-keta-Lambda}
\end{align}
Here, we introduce $\eta_0$ as the time when the tensor modes got excited from the Bunch-Davies vacuum state.
Then, if we properly choose $\eta_0$ so that 
$1 < -k_S\eta_0\leq O(10)$ is satisfied, the aforementioned issues can be absent even in the case where the short-wavelength tensor modes are on the subhorizon scales at $\eta=\eta_0$.

It has been extensively discussed that the non-Bunch-Davies tensor modes can enhance primordial bispectra compared to those in the case of the Bunch-Davies state. In particular, the enhancements of the so-called squeezed bispectra are in proportion to powers of $k_S/k_L$~(see, e.g. Ref.~\cite{Ganc:2012ae}). However, based on the above discussion, we would like to stress that this enhancement appears in the case with Eq.~(\ref{eq:ineq-TransPlanckian}), whereas it does not with Eq.~(\ref{eq:new-ineq-TransPlanckian}).

\subsection{STT bispectrum}

The STT bispectrum is defined by
\begin{align}
\langle\hat\zeta({\bf k}_1)\hat\xi^{(s_2)}({\bf k}_2)\hat\xi^{(s_3)}({\bf k}_3)\rangle=(2\pi)^3\delta({\bf k}_1+{\bf k}_2+{\bf k}_3)\mathcal{B}_{shh}^{s_2s_3}.
\label{eq:crossbis}
\end{align}
By expanding the action Eq.~(\ref{eq: Horndeski action}) up to the cubic order in the perturbations, the interaction Hamiltonian involving one scalar and two tensors is obtained as~\cite{Gao:2012ib}:
\begin{align}
H_{\rm int}=-\int{\rm d}^3x\mathcal{L}_{shh},
\end{align}
with
\begin{align}
\mathcal{L}_{shh}&=a^3\biggl[b_1\zeta\dot h_{ij}^2+\frac{b_2}{a^2}\zeta(\partial_k h_{ij})^2+b_3(\partial_k\psi)\dot h_{ij}\partial_k h_{ij} \nonumber
\\
&\qquad \qquad +b_4\dot\zeta\dot h_{ij}^2+\frac{b_5}{a^2}\partial^2\zeta\dot h_{ij}^2+b_6\psi_{,ij}\dot h_{ik}\dot h_{jk}+\frac{b_7}{a^2}\zeta_{,ij}\dot h_{ik}\dot h_{jk}\biggr]. \label{eq: cubic-lagrangian}
\end{align}
As is summarized in Appendix~\ref{App: perturbed-action}, the above cubic Lagrangian includes other terms proportional to the equations of motion for the linear perturbations which, in principle, can contribute to the bispectrum. However, those are always subleading under the slow-roll approximations, and hence we do not consider those in the main text. Also, throughout the present paper, we approximate all coefficients of cubic interactions to constants under the slow-roll approximations.

For later convenience, we parametrize the primordial bispectrum as
\begin{align}
\mathcal{B}_{shh}^{s_2s_3}&={\rm Re}\left[\sum^{7}_{i=1}\left(\alpha^{(s_2)}_{k_2}-\beta^{(s_2)}_{k_2}\right)\left(\alpha^{(s_3)}_{k_3}-\beta^{(s_3)}_{k_3}\right)\frac{2}{k_1^3k_2^3k_3^3}\frac{H^6}{\mathcal{F}_S\mathcal{F}_T^2c_sc_h^2}b_i\mathcal{F}_{b_i}\mathcal{V}_{b_i}+(s_2, k_2\leftrightarrow s_3, k_3)\right].
\label{eq:Bshh}
\end{align}
The functions $\mathcal{V}_{b_i}$ originate from products of the polarization tensors, the explicit expressions of which are summarized in Appendix~\ref{App: cross-bispectrum}. 
To compute the primordial bispectrum, we perform the time integral of the in-in formalism from $\eta=\eta_0$ to the end of inflation, $\eta=0$. In general, the time integral can be schematically expressed as
\begin{align}
S(\tilde K):=\int_{\eta_0}^0\D\eta (-\eta)^n e^{-i\tilde K\eta},
\end{align}
where $\tilde K$ takes $K', K_1', K_2'$, or $K_3'$ and they are respectively defined as
\begin{align}
K'&:=c_sk_1+c_h(k_2+k_3),\\
K_1'&:=-c_sk_1+c_h(k_2+k_3),\\
K_2'&:=c_sk_1+c_h(-k_2+k_3),\\
K_3'&:=c_sk_1+c_h(k_2-k_3).
\end{align}
Under Eq.~(\ref{eq:new-ineq-TransPlanckian}) with $k_1 = k_L$ and $k_2\simeq k_3 \simeq k_S$, one can obtain
\begin{align}
S(\tilde K)
=\begin{cases}
\displaystyle\mathcal{O}(\tilde K^{-(n+1)})
 &\ \ (\tilde K=K'\ {\rm or}\ K_1'),\\
\displaystyle\mathcal{O}(\eta_0^{(n+1)})
 &\ \ (\tilde K=K_2'\ {\rm or}\ K_3').
\end{cases}
\end{align}
It is worth mentioning the difference of $S(\tilde K)$ between Eq.~(\ref{eq:ineq-TransPlanckian}) and Eq.~(\ref{eq:new-ineq-TransPlanckian}). In the former, one has $S(\tilde K)=\mathcal{O}(\tilde K^{-(n+1)})$ for all four cases under the squeezed limit, which has resulted in the enhancements of the squeezed bispectrum being in proportion to powers of $k_S/k_L$. 
In the latter, one has $S(\tilde K)=\mathcal{O}( \eta_0^{(n+1)})$, which will result in the enhancements being in proportion to powers of $-k_S\eta_0$ as will be shown below. Since we have $k_S/k_L\gg -k_S\eta_0$, the enhancements will be reduced compared to the results in the literature (e.g. Ref.~\cite{Ganc:2012ae}) where Eq.~(\ref{eq:ineq-TransPlanckian}) was assumed. 

We summarize the computational details in Appendix~\ref{App: cross-bispectrum}, and we here show only the final results. The primordial STT bispectrum obtained under Eq.~(\ref{eq:new-ineq-TransPlanckian}) is as follows,
\begin{align}
\mathcal{F}_{b_1}&=\frac{c_h^4k_2^2k_3^2}{H^2}\biggl[\mathcal{I}_{b_1,K'}\alpha^{(s_2)*}_{k_2}\alpha^{(s_3)*}_{k_3}+\mathcal{I}^*_{b_1,K_1'}\beta^{(s_2)*}_{k_2}\beta^{(s_3)*}_{k_3}+\mathcal{I}_{b_1,K_2'}\beta^{(s_2)*}_{k_2}\alpha^{(s_3)*}_{k_3}+\mathcal{I}_{b_1,K_3'}\alpha^{(s_2)*}_{k_2}\beta^{(s_3)*}_{k_3}\biggr],\\
\mathcal{F}_{b_2}&=\frac{1}{H^2}\biggl[\mathcal{I}_{b_2,K'}\alpha^{(s_2)*}_{k_2}\alpha^{(s_3)*}_{k_3}+\mathcal{I}^*_{b_2,K_1'}\beta^{(s_2)*}_{k_2}\beta^{(s_3)*}_{k_3}+\mathcal{I}_{b_2,K_2'}\beta^{(s_2)*}_{k_2}\alpha^{(s_3)*}_{k_3}+\mathcal{I}_{b_2,K_3'}\alpha^{(s_2)*}_{k_2}\beta^{(s_3)*}_{k_3}\biggr],\\
\mathcal{F}_{b_3}&=\frac{c_h^2c_s^2k_1^2k_2^2}{H^2}\biggl[\mathcal{I}_{b_3,K'}\alpha^{(s_2)*}_{k_2}\alpha^{(s_3)*}_{k_3}+\mathcal{I}^*_{b_3,K_1'}\beta^{(s_2)*}_{k_2}\beta^{(s_3)*}_{k_3}+\mathcal{I}_{b_3,K_2'}\beta^{(s_2)*}_{k_2}\alpha^{(s_3)*}_{k_3}+\mathcal{I}_{b_3,K_3'}\alpha^{(s_2)*}_{k_2}\beta^{(s_3)*}_{k_3}\biggr],\\
\mathcal{F}_{b_4}&=\frac{c_h^4c_s^2k_1^2k_2^2k_3^2}{H}\biggl[\mathcal{I}_{b_4,K'}\alpha^{(s_2)*}_{k_2}\alpha^{(s_3)*}_{k_3}+\mathcal{I}^*_{b_4,K_1'}\beta^{(s_2)*}_{k_2}\beta^{(s_3)*}_{k_3}+\mathcal{I}_{b_4,K_2'}\beta^{(s_2)*}_{k_2}\alpha^{(s_3)*}_{k_3}+\mathcal{I}_{b_4,K_3'}\alpha^{(s_2)*}_{k_2}\beta^{(s_3)*}_{k_3}\biggr],\\
\mathcal{F}_{b_5}&=c_h^4k_1^2k_2^2k_3^2\biggl[\mathcal{I}_{b_5,K'}\alpha^{(s_2)*}_{k_2}\alpha^{(s_3)*}_{k_3}+\mathcal{I}^*_{b_5,K_1'}\beta^{(s_2)*}_{k_2}\beta^{(s_3)*}_{k_3}+\mathcal{I}_{b_5,K_2'}\beta^{(s_2)*}_{k_2}\alpha^{(s_3)*}_{k_3}+\mathcal{I}_{b_5,K_3'}\alpha^{(s_2)*}_{k_2}\beta^{(s_3)*}_{k_3}\biggr],\\
\mathcal{F}_{b_6}&=\mathcal{F}_{b_4},\\
\mathcal{F}_{b_7}&=\mathcal{F}_{b_5},
\end{align}
where
\begin{align}
\mathcal{I}_{b_i,K'}&=\mathcal{J}_{b_i}(k_1,k_2,k_3), \label{eq: Ikb1}\\
\mathcal{I}_{b_i,K_1'}&=-\mathcal{J}_{b_i}(-k_1,k_2,k_3), \label{eq: Ik1b1}\\
\mathcal{I}_{b_i,K_2'}&=-\mathcal{J}_{b_i}(k_1,-k_2,k_3)+\mathcal{S}_{b_i}(k_1,-k_2,k_3),\\
\mathcal{I}_{b_i,K_3'}&=-\mathcal{J}_{b_i}(k_1,k_2,-k_3)+\mathcal{S}_{b_i}(k_1,k_2,-k_3),
\end{align}
with
\begin{align}
\mathcal{J}_{b_1}(k_1,k_2,k_3)&:=\frac{c_sk_1+K'}{K'^2},\\
\mathcal{S}_{b_1}(k_1,k_2,k_3)&:=\frac{e^{iK'\eta_0}}{K'^2}(c_sk_1+K'-ic_sk_1K'\eta_0),\\
\mathcal{J}_{b_2}(k_1,k_2,k_3)&:=- \frac{1}{K'^2}\biggl[c_s^3k_1^3+2c_s^2c_hk_1^2(k_2+k_3) \nonumber \\
&\qquad\qquad \quad +2c_sc_h^2k_1(k_2^2+k_2k_3+k_3^2)+c_h^3(k_2+k_3)(k_2^2+k_2k_3+k_3^2)\biggr],\\
\mathcal{S}_{b_2}(k_1,k_2,k_3)&:=e^{iK'\eta_0}\biggl[\frac{c_h}{K'^2}(c_h^2k_2k_3(k_2+k_3)+c_s^2k_1^2(k_2+k_3)
\nonumber \\
& \qquad \qquad \quad +c_sc_hk_1(k_2^2+4k_2k_3+k_3^2))+\frac{i}{\eta_0}-\frac{ic_sc_h^2k_1k_2k_3\eta_0}{K'}\biggr],\\
\mathcal{J}_{b_3}(k_1,k_2,k_3)&:=\frac{K'+c_hk_3}{K'^2},\\
\mathcal{S}_{b_3}(k_1,k_2,k_3)&:=\frac{e^{iK'\eta_0}}{K'^2}(K'+c_hk_3-ic_hK'k_3\eta_0),\\
\mathcal{J}_{b_4}(k_1,k_2,k_3)&:=\frac{2}{K'^3},\\
\mathcal{S}_{b_4}(k_1,k_2,k_3)&:=-\frac{i}{K'^3}e^{iK'\eta_0}(2i+2K'\eta_0-iK'^2\eta_0^2),\\
\mathcal{J}_{b_5}(k_1,k_2,k_3)&:=\frac{2(3c_sk_1+K')}{K'^4},\\
\mathcal{S}_{b_5}(k_1,k_2,k_3)&:=e^{iK'\eta_0}\biggl\{\frac{2(3c_sk_1+K')}{K'^4}
\nonumber \\
& \qquad \qquad \quad
+\frac{i\eta_0}{K'^3}[K'(-2+iK'\eta_0)+c_sk_1(-6+3iK'\eta_0+K'^2\eta_0^2)]\biggr\}.
\end{align}
For the Bunch-Davies state limit ($\beta^{(s)}_{k} \to 0$), this expression gives the previous result in Ref.~\cite{Gao:2012ib}. In particular, in the squeezed limit 
the difference from the case with the Bunch-Davies state
appears in $\mathcal{I}_{b_i,K'_{(2,3)}}$, and 
$\mathcal{I}_{b_i,K'_{(2,3)}}/\mathcal{I}_{b_i,K'}$ corresponds to the characteristic scale-dependent enhancement of the squeezed bispectrum with the non-Bunch-Davies states. By taking the squeezed limit
($k_L = k_1 \ll k_2 \simeq k_3 \simeq k_S$),
we have
\begin{align}
K'\simeq K_1'\simeq2c_hk_S, K_2'\simeq K_3'\simeq c_sk_L, \label{eq: Kprime-squeezed}
\end{align}
and $\mathcal{I}_{b_i,\tilde K}$ can be simply estimated as
\begin{align}
\mathcal{I}_{b_1,K'}&\simeq-\mathcal{I}_{b_1,K_1'}\simeq\frac{1}{2c_hk_S},\ \mathcal{I}_{b_1,K_2'}\simeq\mathcal{I}_{b_1,K_3'}\simeq i\eta_0,\\
\mathcal{I}_{b_2,K'}&\simeq-\mathcal{I}_{b_2,K_1'}\simeq\frac{3c_hk_S}{2},\ \mathcal{I}_{b_2,K_2'}\simeq\mathcal{I}_{b_2,K_3'}\simeq -ic_h^2k_S^2\eta_0,\\
\mathcal{I}_{b_3,K'}&\simeq-\mathcal{I}_{b_3,K_1'}\simeq\frac{3}{4c_hk_S},\ \mathcal{I}_{b_3,K_2'}\simeq-\mathcal{I}_{b_3,K_3'}\simeq \frac{c_hk_S}{2}\eta_0^2,\\
\mathcal{I}_{b_4,K'}&\simeq-\mathcal{I}_{b_4,K_1'}\simeq\frac{1}{4c_h^3k_S^3},\ \mathcal{I}_{b_4,K_2'}\simeq\mathcal{I}_{b_4,K_3'}\simeq-\frac{i}{3}\eta_0^3,\\
\mathcal{I}_{b_5,K'}&\simeq-\mathcal{I}_{b_5,K_1'}\simeq\frac{1}{4c_h^3k_S^3},\ \mathcal{I}_{b_5,K_2'}\simeq\mathcal{I}_{b_5,K_3'}\simeq-\frac{i}{3}\eta_0^3.
\end{align}
Accordingly, we have shown explicitly that the STT bispectrum with the non-Bunch-Davies states in the squeezed limit can be enhanced in proportion to powers of $k_S\eta_0$, compared to that with the Bunch-Davies state. In Sec.~\ref{Sec: SGWB}, we will investigate how those enhancements can leave their imprints on the SGWB anisotropies.

\section{SGWB anisotropies}\label{Sec: SGWB}
\subsection{General expression for the anisotropies}\label{sec: general expression}
It has been shown in Refs.~\cite{Jeong:2012df,Dai:2013kra,Bahrami:2013isa,Dimastrogiovanni:2014ina,Dimastrogiovanni:2015pla,Ricciardone:2017kre,Dimastrogiovanni:2018uqy,Dimastrogiovanni:2019bfl,Bartolo:2019oiq,Bartolo:2019yeu,Adshead:2020bji,Malhotra:2020ket,Dimastrogiovanni:2021mfs,Dimastrogiovanni:2022afr,Orlando:2022rih} that the primordial non-Gaussianities result in spatial modulations of the primordial scalar and tensor power spectra. In particular, the STT bispectrum induces 
the directional dependence in the tensor power spectrum:~\cite{Adshead:2020bji,Dimastrogiovanni:2021mfs,Orlando:2022rih}
\begin{align}
\mathcal{P}_{h}^{(s)} (k,\hat{n})=\bar{\mathcal{P}}^{(s)}_h(k)\biggl[1+
\sum_{s'}\int\frac{\D^3 q}{(2\pi)^3}e^{id {\bf q}\cdot \hat{n}} F^{s s'}_{\rm NL}({\bf k},{\bf q})\zeta({\bf q})\biggr],
\end{align}
where $\hat{n}:= -{\bf k}/k$ and 
$d:= \eta_{\rm now} - \eta_{\rm ent}$ with $\eta_{\rm now}$ and $\eta_{\rm ent}$ being the conformal time at present and at the horizon reentry of inflationary GWs for each $k$ mode, respectively. Since we consider the nHz or deci-Hz inflationary GWs that reentered the horizon long before the present time, i.e. $\eta_{\rm now} \gg \eta_{\rm ent}$, we assume $d \simeq \eta_{\rm now}$.
Hereinafter, a bar means the homogeneous component.
In the above expression, $F^{ss'}_{\rm NL}({\bf k}, {\bf q})$ can be characterized by the cross bispectrum~\eqref{eq:crossbis} as
\begin{align}
F^{ss'}_{\rm NL}({\bf k}, {\bf q}):=\mathcal{B}^{ss'}_{shh}({\bf k}-{\bf q}/2,-{\bf k}-{\bf q}/2,{\bf q})|_{q\to0}\biggl(\frac{\pi^2}{k^3}\mathcal{P}^{(s)}_h(k)\biggr)^{-1}\biggl(\frac{2\pi^2}{q^3}\mathcal{P}_\zeta(q)\biggr)^{-1},
\end{align}
with
\begin{align}
{\bf k}_1 = {\bf q},~{\bf k}_2 = {\bf k} - {\bf q}/2,~{\bf k}_3 = - {\bf k} - {\bf q}/2.
\end{align}
Note that $k$ and $q$ in this section corresponds to $k_L(=k_1)$ and $k_S(\simeq k_2\simeq k_3)$ used in the above sections, respectively. 
As for the SGWB,
the density parameter per logarithmic wavenumber,
\begin{align}
\Omega_{\rm GW} := \frac{1}{\rho_{\rm cr}}
\frac{\D \rho_{\rm GW}}{\D \ln k}, 
\end{align}
has been widely used, where $\rho_{\rm cr}$ is the critical density at present, and for the inflationary GWs it is proportional to the primordial tensor power spectrum.
Thus, the anisotropies of the SGWB induced from the primordial STT bispectrum can be characterized as~\cite{Adshead:2020bji,Dimastrogiovanni:2021mfs,Orlando:2022rih}:
\begin{align}
\Omega_{\rm GW}(k,\hat n)=\bar{\Omega}_{\rm GW}(k)\biggl[1+\delta_{\rm GW}(k,\hat n)\biggr],
\end{align}
where
\begin{align}
\delta_{\rm GW}(k,\hat n) := \int\frac{\D^3 q}{(2\pi)^3}e^{id{\bf q}\cdot\hat n}\sum_{s s'} \tilde{F}^{s s'}_{\rm NL}({\bf k},{\bf q})\zeta({\bf q}),
\end{align} 
with
\begin{align}
\tilde{F}_{\rm NL}^{ss'} ({\bf k},{\bf q}) := \frac{\bar{\mathcal{P}}^{(s)}_h F^{s s'}_{\rm NL}({\bf k},{\bf q})} {\sum_s \bar{\mathcal P}^{(s)}_h}.   
\end{align}
One can expand the anisotropies in terms of the spherical harmonics, the expansion coefficients of which are of the form
\begin{align}
\delta^{\rm GW}_{\ell m}(k)=\int\D^2\hat n Y^*_{\ell m}(\hat n)\delta_{\rm GW}(k,\hat n).
\end{align}
In general, the three-point correlation functions are proportional to products of the polarization tensors that include angular dependence and $F^{s_2s_3}_{\rm NL}$ can be decomposed in terms of the spherical harmonics as
\begin{align}
\tilde{F}^{s_2s_3}_{\rm NL}({\bf k},{\bf q})=Y_{00}(\hat n\cdot\hat q)f^{s_2s_3}_{\rm NL,0}+ Y_{20}(\hat n\cdot\hat q)f^{s_2s_3}_{\rm NL,2}+Y_{40}(\hat n\cdot\hat q)f^{s_2s_3}_{\rm NL,4}+\cdots=\sum_{L={\rm even}}f^{s_2s_3}_{{\rm NL},L} Y_{L0}(\hat n\cdot\hat q),
\end{align}
where $\hat q:={\bf q}/q$.
With this parametrization, we have 
\begin{align}
\delta^{\rm GW}_{\ell m} (k)&=4\pi \sum_{L',M,M'}\sum_{L={\rm even}}i^{L'}\sqrt{(2L'+1)(2\ell+1)}
\begin{pmatrix}
\ell & L' & L \\
0 & 0 & 0
\end{pmatrix}
\begin{pmatrix}
\ell & L' & L \\
m & M' & M
\end{pmatrix}\notag\\
&\quad\times\int\frac{\D^3 q}{(2\pi)^3}Y_{L'M'}(\hat q)Y_{L M}(\hat q)j_{L'}(qd)\sum_{s_2,s_3}f^{s_2s_3}_{{\rm NL}, L}(k,q)\zeta_{\bf q},
\end{align}
with $\begin{pmatrix}
l_1 & l_2 & l_3 \\
L_1 & L_2 & L_3
\end{pmatrix}$ and $j_L(x)$ being the Wigner-$3j$ symbol and spherical Bessel function, respectively.

\subsection{Angular correlation for the SGWB anisotropies}

Based on the previous expression, as statistical observables for the SGWB anisotropies, we can introduce the angular auto-correlation function $C^{\rm GW}_\ell$ as
\begin{align}
C^{\rm GW}_\ell(k)=\frac{1}{2\ell+1}\sum_m\langle\delta^{\rm GW}_{\ell m}(k)\delta^{\rm GW*}_{\ell m}(k)\rangle=\sum_{L={\rm even}}C^{\rm GW}_{\ell,L}(k),
\end{align}
where
\begin{align}
C^{\rm GW}_{\ell, L}(k)&= (2 L +1)\sum_{J,J'}i^{J-J'}(2J+1)(2J'+1)
\biggl[\begin{pmatrix}
\ell & J & L \\
0 & 0 & 0
\end{pmatrix}\biggr]^2
\biggl[\begin{pmatrix}
\ell & J' & L \\
0 & 0 & 0
\end{pmatrix}\biggr]^2\notag\\
&\quad \times \int\frac{\D q}{q}j_J(qd) j_{J'}(qd)\mathcal{P}_\zeta(q)\sum_{s_2,s_3,s'_2,s'_3}f^{s_2s_3}_{{\rm NL}, L}(k,q)f^{s'_2s'_3}_{{\rm NL}, L}(k,q). \label{eq: general-cl-gw}
\end{align}

We also consider the cross-correlation between the SGWB and CMB anisotropies.
The CMB anisotropies can be expanded in terms of the spherical harmonics as
\begin{align}
X(\hat n)=\sum_{\ell m}a^X_{\ell m}Y^*_{\ell m}(\hat n),
\end{align}
with
\begin{align}
a^X_{\ell m}=4\pi i^\ell \int\frac{\D^3p}{(2\pi)^3}\mathcal{T}^X_\ell(p)Y^*_{\ell m}(\hat p)\zeta({\bf p}),
\end{align}
where the $X$ stands for $T$ (temperature fluctuations) or $E$ (E-mode polarization).\footnote{As has been studied in Ref.~\cite{Orlando:2022rih}, the cross-correlation between the B-mode polarization and the SGWB anisotropy sourced by the long-wavelength tensor mode exists in parity-violating gravitational theories.} 
In the same way as the angular auto-correlation function of the SGWB anisotropies, the cross-correlation function of the CMB and SGWB anisotropies can be expressed as
\begin{align}
C^{\rm GW-X}_\ell(k) =\frac{1}{2\ell+1}\sum_m \langle\delta^{\rm GW}_{\ell m}(k)a^{X*}_{\ell m}\rangle=\sum_{L={\rm even}}C^{\rm GW-X}_{\ell,L}(k),
\end{align}
with
\begin{align}
C^{\rm GW-X}_{\ell,L}(k)=4\pi \sqrt{\frac{2 L +1}{4\pi}}\sum_J i^{J-\ell}(2J+1)
\biggl[\begin{pmatrix}
\ell & J & L \\
0 & 0 & 0
\end{pmatrix}\biggr]^2\int\frac{\D q}{q}j_J(qd)\mathcal{T}^X_\ell(q)\mathcal{P}_\zeta(q)
\sum_{s_2,s_3}f^{s_2s_3}_{{\rm NL},L}(k,q),
\end{align}
where $\mathcal{T}^X_\ell(q)$ is the transfer function for the CMB anisotropies.

\subsection{SGWB anisotropies from inflation with the non-Bunch-Davies states}

Then, let us evaluate the angular correlation function sourced from the primordial
STT bispectrum with the non-Bunch-Davies states.
In the present setup, one can take the squeezed limit to the STT bispectrum~\eqref{eq:Bshh} straightforwardly as
\begin{align}
\frac{\mathcal{B}_{shh}^{ss}}{P_s(q) P_h^{(s)}(k) }&=4 \left(\frac{b_1}{\mathcal{G}_T} + \frac{3 b_2}{\mathcal{F}_T}\right) {\rm Re} \left[ \frac{(\alpha^{(s)}_{k_2} - \beta^{(s)}_{k_2}) (\alpha^{(s)}_{k_3} - \beta^{(s)}_{k_3}) (\alpha^{(s)*}_{k_2}\alpha^{(s)*}_{k_3} - \beta^{(s)*}_{k_2}\beta^{(s)*}_{k_3})}{|\alpha^{(s)}_k - \beta^{(s)}_k|^2}\right] \notag\\
&\quad 
-8 \left( \frac{b_1}{\mathcal{G}_T} + \frac{b_2}{\mathcal{F}_T} \right) 
{\rm Im} \left[\frac{(\alpha^{(s)}_{k_2} - \beta^{(s)}_{k_2}) (\alpha^{(s)}_{k_3} - \beta^{(s)}_{k_3}) (\beta^{(s)*}_{k_2}\alpha^{(s)*}_{k_3} + \alpha^{(s)*}_{k_2}\beta^{(s)*}_{k_3})}{|\alpha^{(s)}_k - \beta^{(s)}_k|^2}\right] c_h k \eta_0 \notag\\
&\quad 
+\frac{8 b_2}{\mathcal{F}_T} {\rm Im} \left[ \frac{(\alpha^{(s)}_{k_2} - \beta^{(s)}_{k_2}) (\alpha^{(s)}_{k_3} - \beta^{(s)}_{k_3}) (\alpha^{(s)*}_{k_2}\alpha^{(s)*}_{k_3} + \beta^{(s)*}_{k_2}\beta^{(s)*}_{k_3})}{|\alpha^{(s)}_k - \beta^{(s)}_k|^2}\right]  
\frac{1}{c_hk\eta_0} 
\notag\\
&\quad
- \frac{4 c_s c_h b_3}{\mathcal{F}_T} \,{\rm Re} \left[\frac{(\alpha^{(s)}_{k_2} - \beta^{(s)}_{k_2}) (\alpha^{(s)}_{k_3} - \beta^{(s)}_{k_3}) (\beta^{(s)*}_{k_2}\alpha^{(s)*}_{k_3} - \alpha^{(s)*}_{k_2}\beta^{(s)*}_{k_3})}{|\alpha^{(s)}_k - \beta^{(s)}_k|^2}\right]  \nonumber \\
& \qquad \qquad \qquad \qquad \qquad \qquad \qquad \qquad \qquad \times (\hat n\cdot\hat q)  \times c_h k \eta_0 \times c_s q \eta_0 
+ O((q/k)^2). \label{eq: squeezed-bispectrum}
\end{align}
Note that here we assume $s_2 = s_3 = s$.
This is because, as can be seen in the expressions
for $\mathcal{V}_i$ at the squeezed limit studied in Appendix~\ref{App: squeezed-limit-polarization-tensor}, $\mathcal{V}_i$ are suppressed at most in the order of $q/k$ for the case with $s_2 = - s_3 $. 
Thus, in the leading order, not only the power spectrum
but also the bispectrum depend on the choice of the helicity 
only through the helicity dependence of the Bogoliubov coefficients.
In the following subsections, we will investigate the correlation functions of the SGWB anisotropies for the two cases of the Bogoliubov coefficients which are introduced in section~\ref{Sec: Bogoliubov-coeff}.

\subsubsection{Peaked spectrum}

The fourth line in the squeezed bispectrum ~(\ref{eq: squeezed-bispectrum}), which is proportional to $\hat n\cdot\hat q$,
has 
\begin{align}
(\beta^{(s)*}_{k_2}\alpha^{(s)*}_{k_3}-\alpha^{(s)*}_{k_2}\beta^{(s)*}_{k_3}). 
\label{eq: mu-term-squeezedB}
\end{align}
By substituting the assumptions for the Bogoliubov coefficients given by Eqs.~(\ref{eq: bogo-alpha}) and~(\ref{eq: bogo-beta}),
one can find that it is reduced to ${\rm Re}[\beta^{(s)}_{k_2}] - {\rm Re}[\beta^{(s)}_{k_3}]$ and at the leading order in the squeezed limit this should be on the order of $(\hat n\cdot\hat q) q$. Thus, in the squeezed limit the fourth line of Eq.~\eqref{eq: squeezed-bispectrum} is found to be proportional to 
$(\hat n\cdot\hat q)^2 \times (c_s q \eta_0)^2$.
Since we focus on the phase after the long-wavelength mode $q$ crosses the sound horizon, i.e. the phase characterized by $|c_s q\eta_0|\ll1$, there is no enhancement for the term proportional to $(\hat n\cdot\hat q)^2$ in the bispectrum. Accordingly, it is enough to consider the terms proportional to $(\hat n\cdot\hat q)^0$.

From Eqs.~(\ref{eq: bogo-alpha}) and~(\ref{eq: bogo-beta}), up to the leading order in $q/k$, we can also find 
\begin{align}
{\rm Re} & \left[\frac{(\alpha^{(s)}_{k_2} - \beta^{(s)}_{k_2}) (\alpha^{(s)}_{k_3} - \beta^{(s)}_{k_3}) (\alpha^{(s)*}_{k_2}\alpha^{(s)*}_{k_3} - \beta^{(s)*}_{k_2}\beta^{(s)*}_{k_3})}{|\alpha^{(s)}_k - \beta^{(s)}_k|^2}\right] \approx 1,\\
{\rm Im} & \left[ \frac{(\alpha^{(s)}_{k_2} - \beta^{(s)}_{k_2}) (\alpha^{(s)}_{k_3} - \beta^{(s)}_{k_3}) (\beta^{(s)*}_{k_2}\alpha^{(s)*}_{k_3} + \alpha^{(s)*}_{k_2}\beta^{(s)*}_{k_3})}{|\alpha^{(s)}_k - \beta^{(s)}_k|^2}\right] \notag \\
& \qquad \qquad  \qquad \approx
{\rm Im}  \left[ \frac{(\alpha^{(s)}_{k_2} - \beta^{(s)}_{k_2}) (\alpha^{(s)}_{k_3} - \beta^{(s)}_{k_3}) (\alpha^{(s)*}_{k_2}\alpha^{(s)*}_{k_3} + \beta^{(s)*}_{k_2}\beta^{(s)*}_{k_3})}{|\alpha^{(s)}_k - \beta^{(s)}_k|^2}\right]  \notag\\
& \qquad \qquad \qquad \approx \frac{2A^2}{2A^2-2A e^{\ln^2(k/k_*)/\Delta^2}+e^{2\ln^2(k/k_*)/\Delta^2}}=:\mathcal{A}(k,k_*).
\end{align}
For $k \approx k_*$ and $A \gg 1$,
$\mathcal{A}(k,k_*) \approx 1$. One can also find $\mathcal{A}\ll1$ for $A\ll1$ which suppresses terms in the non-linearity parameter newly induced by non-Bunch-Davies effects as shown below. Therefore, $A\gg1$ is found to be important to enhance both SGWBs and anisotropies.
As a result, 
the non-linearity parameter in the squeezed limit of the STT bispectrum $F_{\rm NL}^{s}$ for the peaked tensor power spectrum can be rewritten as
\begin{align}
\label{eq:fNLexpression}
\tilde{F}_{\rm NL}^{ss}:=\frac{\mathcal{B}^{ss}_{shh}|_{q \to 0}}{P_s(q) P_h^{(s)}(k)}&=
\left( \frac{4 b_1}{\mathcal{G}_T} + \frac{12 b_2}{\mathcal{F}_T} \right)
- \left[8 \left(\frac{b_1}{\mathcal{G}_T} + \frac{b_2}{\mathcal{F}_T}  \right) c_h k\eta_0 - \frac{8 b_2}{\mathcal{F}_T} \frac{1}{c_h k\eta_0} \right] \mathcal{A}(k,k_*),
\end{align}
and also 
\begin{align}
f^{ss}_{{\rm NL},0}(k)=\frac{\tilde{F}^{ss}_{\rm NL}}{Y_{00}}=2\sqrt{\pi}\biggl\{\left( \frac{4 b_1}{\mathcal{G}_T} + \frac{12 b_2}{\mathcal{F}_T} \right)
- \left[8 \left(\frac{b_1}{\mathcal{G}_T} + \frac{b_2}{\mathcal{F}_T}  \right) c_h k\eta_0 - \frac{8 b_2}{\mathcal{F}_T} \frac{1}{c_h k\eta_0} \right] \mathcal{A}(k,k_*)\biggr\}. \label{eq: result-fNL}
\end{align}
Note that the leading-order non-linearity parameter is independent of $q$ in the squeeze limit.

\begin{figure} [htb]
     \begin{tabular}{cc}
        \begin{minipage}{0.45\hsize}
            \centering
            \includegraphics[width=7.cm]{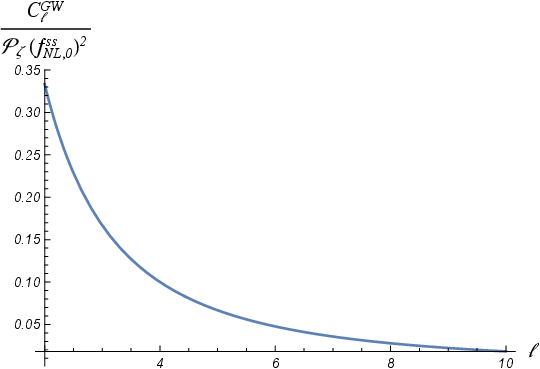}
        \end{minipage} &
        \begin{minipage}{0.45\hsize}
            \centering
            \includegraphics[width=7.cm]{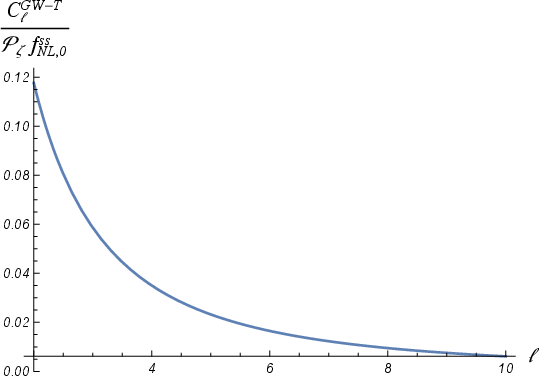}
        \end{minipage} 
    \end{tabular}
    \caption{\emph{Left}: The plot of $C_\ell^{\rm GW}$ normalized by $\mathcal{P}_\zeta f^{ss}_{{\rm NL},0}(k)^2$. \emph{Right}: The plot of $C_\ell^{\rm GW-T}$ normalized by $\mathcal{P}_\zeta f^{ss}_{{\rm NL},0}(k)$.}
    \label{Fig: Cl}
\end{figure}

We then compute the correlation functions of the SGWB anisotropies. From Eq.~(\ref{eq: general-cl-gw}), we obtain
\begin{align}
C^{\rm GW}_{\ell} (k)=\frac{4 \mathcal{P}_\zeta f^{ss}_{{\rm NL},0}(k)^2}{2\ell(\ell+1)},
\label{eq: ClGW}
\end{align}
where we assumed the scale-invariant scalar power spectrum and we used
\begin{align}
\int^{\infty}_{0} \frac{\D x}{x}j_\ell(x)j_\ell(x)= \frac{1}{2\ell(\ell+1)}. \label{eq: integral-self}
\end{align}
For the cross-correlations between the SGWB and CMB anisotropies, we only consider the temperature fluctuation since the amplitude of the E-mode polarization is much smaller than that of the temperature fluctuation. Since we are interested in the temperature fluctuation on a large scale where the SW effect dominates other effects, as the transfer function for the temperature anisotropy we use the following approximate form:
\begin{align}
\mathcal{T}^T_\ell(k)\simeq \frac{1}{5}j_\ell(k(\eta_{\rm now}-\eta_{\rm dec})), \label{eq: approx-transfer}
\end{align}
where $\eta_{\rm dec}$ denotes the decoupling time.
In the calculation of the correlation functions, we use $\eta_{\rm dec}/\eta_{\rm now}\sim 0.02$. As a result, we can rewrite $C_\ell^{\rm GW-T}$ as
\begin{align}
C^{\rm GW-T}_{\ell}(k) \simeq \frac{4\sqrt{\pi}}{5}\mathcal{P}_\zeta 
f^{ss}_{{\rm NL},0}(k)\int\frac{\D q}{q}j_\ell(q d)j_\ell(q(\eta_{\rm now}-\eta_{\rm dec})),
\end{align}
the integral of which can be performed analytically as
\begin{align}
\int_0^\infty\frac{\D q}{q}j_\ell(q d)j_\ell(q(\eta_{\rm now}-\eta_{\rm dec}))&= \int_0^\infty\frac{\D (qd)}{qd}j_\ell(q d)j_\ell(qd(\eta_{\rm now}-\eta_{\rm dec})/d)\notag\\
&=\frac{1}{4}c^\ell\sqrt{\pi}\Gamma(\ell)\frac{{}_{2}F_1\biggl(-\frac{1}{2},\ell,\frac{3}{2}+\ell;c^2\biggr)}{\Gamma(\frac{3}{2}+\ell)}, \label{eq: integral-cross}
\end{align}
where ${}_2F_1(a,b,c;z)$ is the hypergeometrical function, and we introduced
\begin{align}
c:=\frac{\eta_{\rm now}-\eta_{\rm dec}}{d}
\simeq \frac{\eta_{\rm now}-\eta_{\rm dec}}{\eta_{\rm now}}.
\end{align}
Note that by taking $c\to1$ to Eq.~(\ref{eq: integral-cross}), 
we obtain the same result with Eq.~(\ref{eq: integral-self}). 
The resultant form of $C_\ell^{{\rm GW}-T}$ reads
\begin{align}
C_\ell^{\rm GW-T} (k)\simeq \frac{\pi}{5}\mathcal{P}_\zeta f^{ss}_{{\rm NL},0}(k) \times c^\ell\frac{\Gamma(\ell)}{\Gamma(\ell+\frac{3}{2})}{}_{2}F_1\biggl(-\frac{1}{2},\ell,\frac{3}{2}+\ell;c^2\biggr).
\label{eq: ClGWT}
\end{align}
The plots of normalized $C_\ell$s are shown in Fig.~\ref{Fig: Cl}. 

\begin{figure} [htb]
     \begin{tabular}{cc}
        \begin{minipage}{0.45\hsize}
            \centering
            \includegraphics[width=7.cm]{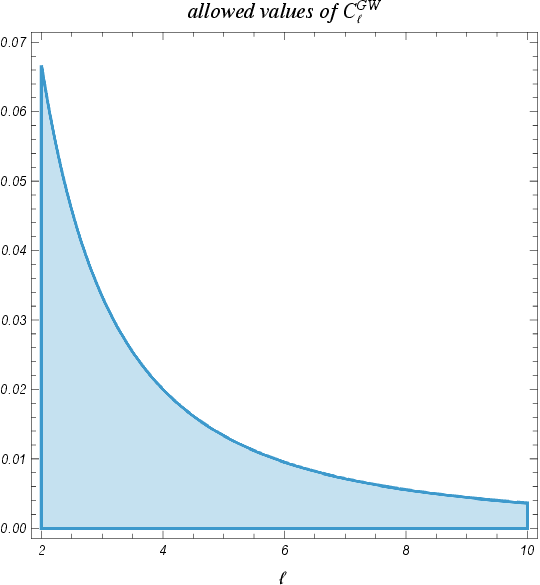}
        \end{minipage} &
        \begin{minipage}{0.45\hsize}
            \centering
            \includegraphics[width=7.cm]{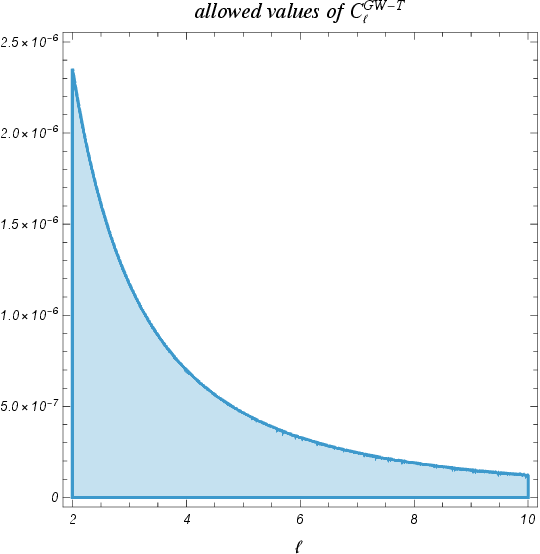}
        \end{minipage} 
    \end{tabular}
    \caption{\emph{Left}: The allowed values of $C_\ell^{\rm GW}$ for the peaked spectrum. \emph{Right}: The allowed values of $C_\ell^{\rm GW-T}$ for the peaked spectrum.}
    \label{Fig: MaxCl}
\end{figure}

Let us investigate the typical amplitudes of the SGWB anisotropies
induced by the primordial STT bispectrum with the non-Bunch-Davies vacuum states. 
To do so, we first need to inspect two theoretical constraints on $b_i$ and $\eta_0$. The first constraint comes from an argument on the validity of the linear perturbation theory. In the present paper, we discuss this point by requiring that the STT interaction is smaller than the quadratic tensor interaction. By requiring $\mathcal{L}^{(2)}_h\geq\mathcal{L}^{(3)}_{shh}$ where $\mathcal{L}^{(2)}_h$ stands for the quadratic Lagrangian of the tensor perturbations, we have
\begin{align}
\frac{b_1}{\mathcal{G}_T}&\leq\mathcal{O}(10^5) \left( \frac{|\zeta(\eta_0)|}{10^{-5}} \right)^{-1}, \label{eq: const-b1}\\
\frac{b_2}{\mathcal{F}_T}&\leq\mathcal{O}(10^5) \left( \frac{|\zeta(\eta_0)|}{10^{-5}} \right)^{-1}. \label{eq: const-b2}
\end{align}
In light of the fact that the curvature perturbation is on the superhorizon scales at $\eta=\eta_0$, we can use the current constraint on the amplitude of $\zeta$ as $|\zeta(\eta_0)| \sim 10^{-5}$~\cite{Planck:2018jri}. 
Another constraint is obtained from an argument on backreaction from the excited tensor modes. Once the modes get excited from the Bunch-Davies state, the excited modes cause backreaction to the inflationary background~\cite{Tanaka:2000jw}. In the Horndeski theory, the excited tensor mode has the following energy density~\cite{Akama:2020jko}:
\begin{align}
\rho_{\rm excited}=\frac{c_h}{a^4(\eta)}\int|\beta^{(s)}_k|^2k^3\D k.
\end{align}
In order for the backreaction from the excited modes not to spoil the background dynamics of the slow-roll inflation, we compare this to the homogeneous part of the Friedmann equation $\mathcal{E}_b$. In the standard slow-roll inflation, the homogeneous part corresponds to the energy density of the inflaton as $\mathcal{E}_b\sim M_{\rm pl}^2 H_{\rm inf}^2$. In the general inflation model considered here, we similarly assume $\mathcal{E}_b\sim M_{\rm pl}^2 H_{\rm inf}^2$ for simplicity. Then, the backreaction constraint, i.e. $\rho_{\rm excited}\lesssim \mathcal{E}_b$, reads
\begin{align}
\frac{1}{a(\eta_0)^4}A^2e^{2\Delta^2}k_*^4\Delta\lesssim M_{\rm pl}^2H_{\rm inf}^2. 
\end{align}
If we assume $A=-3000$, $\Delta=0.45$, and $H_{\rm inf} = 10^{13}~{\rm GeV}$ to explain the recent PTA results, the above reads
\begin{align}
-k_*\eta_0\lesssim 10, \label{eq: const-eta0}
\end{align}
which is consistent with Eq.~(\ref{eq: constraint-keta-Lambda}).
From Eqs.~(\ref{eq: const-b1}),~(\ref{eq: const-b2}), and~(\ref{eq: const-eta0}), we can estimate the theoretical upper bound on $f_{\rm NL}$ as
\begin{align}
f^{ss}_{{\rm NL},0}(k) \lesssim 10^7. \label{eq: pert-constraint-fnl}
\end{align}
Here, we have chosen the parameters in the Bogoliubov coefficients $A$ and $\Delta$ in such a way that the SGWBs can be detected with PTA experiments. As a result, the backreaction constraint has put a constraint on only $-k_*\eta_0$ as in Eq.~(\ref{eq: const-eta0}) that  does not depend on the choice of the cutoff scale.
We should also discuss an observational constraint on $f^{ss}_{{\rm NL},0}(k)$. So far, the CMB temperature bispectrum $B_{TTT}$ (i.e. a non-linearity parameter for a scalar non-Gaussianity) has been roughly constrained as $B_{TTT}/P_\zeta^2\lesssim 10$~\cite{Planck:2019kim}. Since both scalar and tensor perturbations generate the CMB temperature fluctuations (i.e. $\mathcal{B}_{shh}$ contributes to $B_{TTT}$), the above constraint naively applies to the STT bispectrum as
\begin{align}
\mathcal{B}_{shh}|_{k_{\rm CMB}}\sim \mathcal{O}\biggl(\frac{b_1}{\mathcal{G}_T},\frac{b_2}{\mathcal{F}_T}\biggr)\times P_\zeta|_{k_{\rm CMB}} P_h|_{k_{\rm CMB}}\lesssim 10P^2_\zeta|_{k_{\rm CMB}},
\end{align}
which yields the following constraint:
\begin{align}
\mathcal{O}\biggl(\frac{b_1}{\mathcal{G}_T},\frac{b_2}{\mathcal{F}_T}\biggr)\lesssim 10^3 \, \left(\frac{r|_{k_{\rm CMB}}}{10^{-2}}\right)^{-1}. 
\label{eq: bispectrumconst}
\end{align}
For $r|_{k_{\rm CMB}}=10^{-2}$, 
we obtain a stronger constraint on $f^{ss}_{{\rm NL},0}(k)$ (i.e. $f^{ss}_{{\rm NL},0}(k) \lesssim 10^4$) than the theoretical one in Eq.~(\ref{eq: pert-constraint-fnl}).
Finally, we can estimate the maximum possible values of $C^{\rm GW}_\ell$ and $C^{\rm GW-T}_\ell$. Since we have assumed the Bunch-Davies scalar mode, we have $\mathcal{P}_\zeta\simeq 2\times 10^{-9}$. Also, the upper bound on $f^{ss}_{{\rm NL},0}(k)$ is evaluated as $10^4$ as discussed above. As a result, we can find that $C^{\rm GW}_\ell$ and $C^{\rm GW-T}_\ell$ can reach $10^{-2}$ and $10^{-6}$, respectively. The allowed values of the correlation functions are shown in Fig.~\ref{Fig: MaxCl}.

Let us mention the possibility of obtaining a large value of $b_{1,2}$ in the framework of slow-roll inflation. So far, we have shown that $f^{ss}_{{\rm NL},0}(k)$ can be enhanced by a factor $-k\eta_0 \, (\lesssim10)$ due to the consequence of the non-Bunch-Davies effects. The largest enhancement of $f^{ss}_{{\rm NL},0}(k)$ can thus arise from large values of the coupling functions of the cubic interactions $b_{1,2}$. Thus, one might expect that similar enhancements could be predicted from inflation with the Bunch-Davies state. However, in that case, the squeezed bispectrum is expected to satisfy the so-called consistency relation between the spectral index $n_t$ and the amplitude of the squeezed bispectrum~\cite{Maldacena:2002vr,Kundu:2013gha}. To show it explicitly in the Horndeski theory, one would need to take into account slow-roll corrections to the scale factor and the mode functions, as has been studied in the context of the scalar non-Gaussianity in Ref.~\cite{DeFelice:2013ar}. Without taking those into account, one can show at least that the squeezed bispectrum with the Bunch-Davies state is suppressed by the slow-roll parameters. Now the terms in the case of the Bunch-Davies state correspond to the ones irrelevant to $k\eta_0$ in Eq.~(\ref{eq: result-fNL}). The explicit forms of $b_1$ and $b_2$ are obtained as
\begin{align}
b_1&=\frac{3\mathcal{G}_T}{8}\biggl\{1-\frac{H\mathcal{G}_T^2}{\Theta\mathcal{F}_T}+\frac{\mathcal{G}_T}{3}\biggl[-\frac{\dot{\mathcal{F}}_T}{\mathcal{F}_T^2}\frac{\mathcal{G}_T}{\Theta}+\frac{1}{\mathcal{F}_T}\frac{\D}{\D t}\biggl(\frac{\mathcal{G}_T}{\Theta}\biggr)\biggr]\biggr\},\\
b_2&=\frac{1}{8}\frac{H\mathcal{G}_T^2}{\Theta}+\frac{1}{8}\frac{\mathcal{G}_T}{\Theta}\dot{\mathcal{G}}_T+\frac{\mathcal{G}_T}{8}\frac{\D}{\D t}\biggl(\frac{\mathcal{G}_T}{\Theta}\biggr)-\frac{\mathcal{F}_T}{8}.
\end{align}
The terms without the time derivatives cancel with each other only in the following combination:
\begin{align}
\frac{b_1}{\mathcal{G}_T}+\frac{3b_2}{\mathcal{F}_T}&=\frac{\sigma}{8c_h^2}\biggl(2\epsilon-f_T+3g_T+\frac{2\dot\sigma}{H}\biggr),
\end{align}
with $\sigma:=H\mathcal{G}_T/\Theta$. Note that $\sigma=1$ in the k-essence theory. The above is nothing but the combination appearing in the case of the Bunch-Davies state and is suppressed by the slow-roll parameters. However, for different combinations (e.g. $b_1/\mathcal{G}_T+b_2/\mathcal{F}_T$), there are terms that are not proportional to the slow-roll parameters. To see more details about the enhancement of $f^{ss}_{{\rm NL},0}(k)$, let us express our $f^{ss}_{{\rm NL},0}(k)$ in terms of $f_{\rm NL,BD}$ which is the result in the case of the Bunch-Davies state as
\begin{align}
f^{ss}_{{\rm NL},0}(k)=f_{\rm NL,BD}-2c_hk\eta_0 \mathcal{A}(k,k_*)\biggl(f_{\rm NL,BD}-\sqrt{\pi}\frac{b_2}{\mathcal{F}_T}\frac{1+2c_h^2k^2\eta_0^2}{c_h^2k^2\eta_0^2}\biggr),
\end{align}
where
\begin{align}
f_{\rm NL,BD}:=8\sqrt{\pi}\biggl(\frac{b_1}{\mathcal{G}_T}+\frac{3b_2}{\mathcal{F}_T}\biggr).
\end{align}
If both scalar and tensor power spectra are almost scale invariant on the CMB scales, we have
\begin{align}
r|_{k_{\rm CMB}}\simeq 16\frac{\mathcal{F}_S}{\mathcal{F}_T}\frac{c_s}{c_h}.
\end{align}
Taking this into account and using $b_2=\mathcal{F}_S/8$, we can rewrite $f^{ss}_{{\rm NL},0}(k)$ as
\begin{align}
f^{ss}_{{\rm NL},0}(k)=f_{\rm NL,BD}-2c_hk\eta_0 \mathcal{A}(k,k_*)\biggl(f_{\rm NL,BD}-\frac{\sqrt{\pi}}{128}\frac{c_h}{c_s}r|_{k_{\rm CMB}}\frac{1+2c_h^2k^2\eta_0^2}{c_h^2k^2\eta_0^2}\biggr).
\end{align}
We can thus estimate the deviation of $f_{\rm NL}$ from $f_{\rm NL,BD}$ as
\begin{align}
|f^{ss}_{{\rm NL},0}(k)-f_{\rm NL,BD}|\leq\mathcal{O}\biggl(\frac{c_h}{c_s}r|_{k_{\rm CMB}}\biggr)\leq \mathcal{O}\biggl(10^{-2}\frac{c_h}{c_s}\biggr), \label{eq: const_fnlnBD_fnlBD}
\end{align}
where we assumed $r|_{k_{\rm CMB}}\sim10^{-2}$ in the last step, which indicates that one needs a small $c_s$. This is dangerous at least within the k-essence theory where a small $c_s$ always indicates a large scalar non-Gaussianity~\cite{Chen:2006nt}. As will be given in Appendix~\ref{App: SGWBanisotropies-and-scalarnon-Gaussianity}, both the maximum enhancement of the SGWB anisotropies and the generation of a small scalar non-Gaussianity are simultaneously achievable in the full Horndeski theory i.e. in the presence of the $G_4/G_5$ terms. Therefore, we conclude that inflation with the non-Bunch-Davies states in the full Horndeski action can potentially enhance $f^{ss}_{{\rm NL},0}(k)$ up to $10^4$ and the SGWB anisotropies $C_\ell^{\rm GW}$ and $C_\ell^{\rm GW-T}$ up to $10^{-2}$ and $10^{-6}$, respectively.

Before moving to the scale-invariant case, let us compare the amplitudes of the SGWB anisotropies from inflation with the non-Bunch-Davies states to those originating from the propagation of the GWs at a late time. On large scales, the dominant contributions to the anisotropies originate from the SW effect as in the CMB fluctuations; the auto- and cross-correlation functions originating from which are, respectively, of the form~\cite{Dimastrogiovanni:2021mfs},
\begin{align}
C_\ell^{\rm GW, SW}&=\frac{128\pi}{9\ell(\ell+1)}\mathcal{P}_\zeta \simeq 3 \times 10^{-9} \left[\frac{30}{\ell (\ell + 1)}\right] \left( \frac{\mathcal{P}_\zeta}{2 \times 10^{-9}} \right),\\
C_\ell^{\rm GW-T, SW}&=\frac{16\pi}{15\ell(\ell+1)}\mathcal{P}_\zeta \simeq 2 \times 10^{-10} \left[ \frac{30}{\ell (\ell + 1)}\right]\left( \frac{\mathcal{P}_\zeta}{2 \times 10^{-9}} \right).
\end{align}
Due to the enhancements by $f^{ss}_{{\rm NL},0}(k)$, the anisotropies originating from inflation with the non-Bunch-Davies state can dominate those from the SW effect.

\subsubsection{Nearly scale-invariant spectrum}
\label{sec:bis_scale_inv}

In the case of the $k$-independent Bogoliubov coefficients,\footnote{In the scale-invariant case, the contribution which is proportional to $\hat n\cdot \hat q$ in the STT bispectrum is zero in the squeezed limit.} we have
\begin{align}
{\rm Re} & \left[\frac{(\alpha^{(s)}_{k_2} - \beta^{(s)}_{k_2}) (\alpha^{(s)}_{k_3} - \beta^{(s)}_{k_3}) (\alpha^{(s)*}_{k_2}\alpha^{(s)*}_{k_3} - \beta^{(s)*}_{k_2}\beta^{(s)*}_{k_3})}{|\alpha^{(s)}_k - \beta^{(s)}_k|^2}\right] = 1, \label{eq: block-bogoliubov-1}\\
{\rm Im} & \left[ \frac{(\alpha^{(s)}_{k_2} - \beta^{(s)}_{k_2}) (\alpha^{(s)}_{k_3} - \beta^{(s)}_{k_3}) (\beta^{(s)*}_{k_2}\alpha^{(s)*}_{k_3} + \alpha^{(s)*}_{k_2}\beta^{(s)*}_{k_3})}{|\alpha^{(s)}_k - \beta^{(s)}_k|^2}\right]  \notag \\
& \qquad \qquad  \qquad
=
{\rm Im} \left[ \frac{(\alpha^{(s)}_{k_2} - \beta^{(s)}_{k_2}) (\alpha^{(s)}_{k_3} - \beta^{(s)}_{k_3}) (\alpha^{(s)*}_{k_2}\alpha^{(s)*}_{k_3} + \beta^{(s)*}_{k_2}\beta^{(s)*}_{k_3})}{|\alpha^{(s)}_k - \beta^{(s)}_k|^2}\right] \notag \\
& \qquad \qquad  \qquad = -2B. \label{eq: block-bogoliubov-2}
\end{align}
Then, $f^{ss}_{{\rm NL},0}(k)$ reads
\begin{align}
f^{ss}_{{\rm NL},0}(k)= 2\sqrt{\pi}\biggl\{\left( \frac{4 b_1}{\mathcal{G}_T} + \frac{12 b_2}{\mathcal{F}_T} \right)
+2B \left[8 \left(\frac{b_1}{\mathcal{G}_T} + \frac{b_2}{\mathcal{F}_T}  \right) c_h k\eta_0 - \frac{8 b_2}{\mathcal{F}_T} \frac{1}{c_h k\eta_0} \right] \biggr\}. \label{eq: result-fNL-2}
\end{align}
Since the above $f^{ss}_{{\rm NL},0}(k)$ is independent of $q$, %
calculations of the correlation functions can be similarly performed as in the previous case with peaked power spectrum, and $C_\ell^{\rm GW}$ and $C_\ell^{\rm GW-T}$ are given by Eqs.~\eqref{eq: ClGW} and~\eqref{eq: ClGWT}, respectively.

As for the constraints on the non-linearity parameters,
the constraints on the coupling functions of the cubic interactions $b_{1,2}$ are also expected to be roughly given by
Eqs.~\eqref{eq: const-b1}, \eqref{eq: const-b2}, and \eqref{eq: bispectrumconst}.
On the other hand, the theoretical backreaction constraint reads
\begin{align}
c_h B^2 \Lambda^4\lesssim \mpl^2 H_{\rm inf}^2,
\end{align}
where we have imposed $k/a(\eta_0)\leq\Lambda$. Note here that the backreaction constraint becomes weaker for a lower cutoff scale, which is in contrast to the peaked-spectrum case in which the backreaction constraint is insensitive to the choice of the cutoff scale. In particular, the parameter $B$ can increase up to $\mathcal{O}(10^5)$ for the  cutoff scale given by Eq.~\eqref{eq: cutoff}. 
By combining the above with Eq.~(\ref{eq: constraint-keta-cutoff}), Eq.~(\ref{eq: result-fNL-2}), and $c_h=\mathcal{O}(1)$, we obtain
\begin{align}
f^{ss}_{\rm NL,0}(k)\sim B(-k\eta_0)\mathcal{O}\left(\frac{b_1}{\mathcal{G}_T},\frac{b_2}{\mathcal{F}_T}\right)\leq \frac{\mpl}{\Lambda}\mathcal{O}(10^3),
\end{align}
which holds for any $\Lambda$.
One can find from the above that as the cutoff scale is lower (i.e. the backreaction constraint is weaker), the maximum possible magnitude of $f^{ss}_{\rm NL,0}$ becomes greater.
In particular, for the cutoff scale given by Eq.~\eqref{eq: cutoff}, we have $f^{ss}_{\rm NL,0}(k)\leq\mathcal{O}(10^7)$ where $f^{ss}_{\rm NL,0}(k)$ can be $\mathcal{O}(10^7)$ for the case of $k/a(\eta_0)\sim\Lambda=(\sqrt{\epsilon}\mpl H_{\rm inf}^2)^{1/3}$. As a result, $C^{\rm GW}_\ell$ and $C^{\rm GW-T}_\ell$ can reach $10^{4}$ and $10^{-3}$, respectively. 
In fact, in this case, since the parameter $B$ can take $O(10^5)$, it can be considered that the SGWB anisotropies are enhanced by the non-Bunch-Davies states. 
Note that the violation of the non-Gaussianity consistency relation still plays an important role in enhancing the anisotropies up to the maximum possible amplitudes.

\begin{figure} [htb]
     \begin{tabular}{cc}
        \begin{minipage}{0.45\hsize}
            \centering
            \includegraphics[width=7.cm]{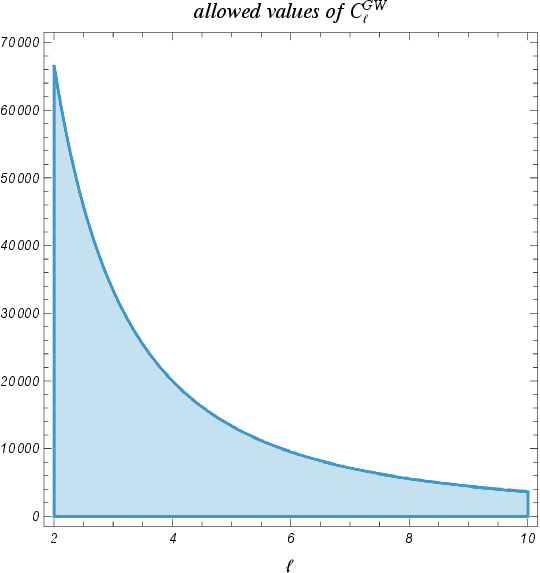}
        \end{minipage} &
        \begin{minipage}{0.45\hsize}
            \centering
            \includegraphics[width=7.cm]{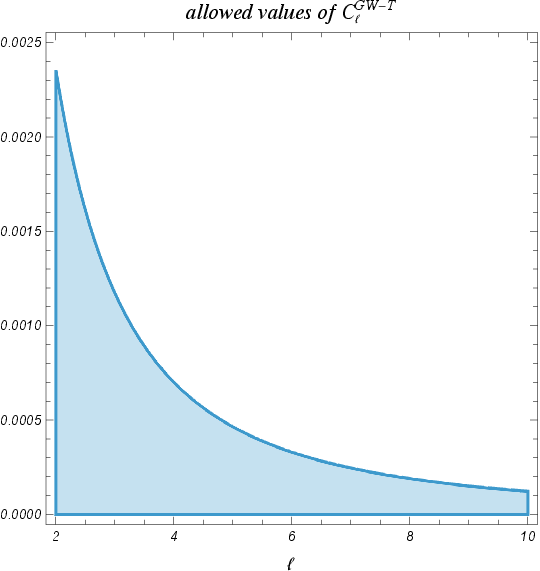}
        \end{minipage} 
    \end{tabular}
    \caption{\emph{Left}: The allowed values of $C_\ell^{\rm GW}$ for the scale-invariant spectrum. \emph{Right}: The allowed values of $C_\ell^{\rm GW-T}$ for the scale-invariant spectrum.}
    \label{Fig: MaxCl2}
\end{figure}

The allowed values of the correlation functions are shown in Fig.~\ref{Fig: MaxCl2}. The amplitudes of GW anisotropies can be greater in the scale-invariant case than in the peaked-spectrum case in which the maximum value of $f^{ss}_{\rm NL,0}(k)$ is $\mathcal{O}(10^4)$.

\subsection{Detectability of the SGWB anisotropies} 
Based on the formulae shown in the above sections for the correlation functions of the SGWB anisotropies
induced by the inflationary STT bispectrum with the non-Bunch-Davies states,
let us discuss the detectability of such inflationary SGWB anisotropies in future experiments. The detectability of the SGWB anisotropies has been discussed for the space-based interferometers BBO/DECIGO and Square Kilometre Array (SKA) as a PTA experiment in Ref.~\cite{Dimastrogiovanni:2021mfs}, where a noise angular spectrum $N_\ell$ for each experiment has been computed for the case of a blue-tilted tensor spectrum. 
In fact, in our paper, we have considered the peaked tensor power spectrum
with the non-Bunch-Davies state to explain the recent SGWB data reported by the PTA experiments such as NANOGrav, and it is not exactly the same as the one assumed in Ref.~\cite{Dimastrogiovanni:2021mfs}.
However, as discussed in the previous section~\ref{sec:intro peaked},
by setting the model parameters appropriately, our model can reproduce the blue-tilted tensor power spectrum for the relevant frequency range of the PTA experiments.
Thus, 
we employ their result of the noise spectrum when discussing the detectability with future PTA experiments such as SKA.
For the scale-invariant case discussed in Sec.~\ref{sec:bis_scale_inv},
we employ the noise spectrum computed
in Ref.~\cite{Ricciardone:2021kel}. We investigate the detectability of the SGWB anisotropies for the case of the scale-independent and -dependent Bogoliubov coefficients with BBO/DECIGO and SKA, respectively.

Since the power spectra of the SGWB anisotropies induced from the primordial STT bispectrum with the non-Bunch-Davies states decrease as the multipole moment $\ell$ increases as shown in the previous section, and also the amplitudes of the noise spectra rapidly increase at $\ell > 10$~\cite{Dimastrogiovanni:2021mfs, Ricciardone:2021kel}, in the estimation of the detectability we only consider the low-$
\ell$ region as $\ell\leq10$.
Following Refs.~\cite{Dimastrogiovanni:2021mfs, Ricciardone:2021kel}, for $\ell\leq10$, 
the typical amplitudes of the noise spectrum are given as $N_\ell = \mathcal{O}(10^{-30})$ in the case of the scale-invariant spectrum for BBO/DECIGO and $N_\ell = \mathcal{O}(10^{-26})$ in the case of the blue-tilted spectrum for SKA, respectively. The noise used for computing the signal-to-noise ratio (SNR) of the SGWB anisotropies is 
\begin{align}
N_\ell^{\rm GW}:=\frac{N_\ell}{\bar\Omega^2_{\rm GW}}.
\end{align}

For the auto-correlation function, we inspect $C^{\rm GW}_\ell/N_\ell^{\rm GW}$ as the SNR,\footnote{Exactly speaking, the SNR squared is evaluated by computing~\cite{LISACosmologyWorkingGroup:2022jok,Orlando:2022rih}:
\begin{align}
\sum_\ell T_{\rm obs}\int\D f (2\ell+1)\frac{C_\ell^{\rm GW}}{N_\ell^{\rm GW}},
\end{align}
where $T_{\rm obs}$ is the total time for observations. In the present paper, we use $C^{\rm GW}_\ell/N_\ell^{\rm GW}$ for the purpose of order estimation.
} i.e. the threshold for detection is roughly estimated as $C^{\rm GW}_\ell/N_\ell^{\rm GW} \gtrsim 1$.
In our case where the cutoff scale is given by Eq.~\eqref{eq: cutoff}, $C_\ell^{\rm GW}$ can increase up to $\mathcal{O}(10^{4})$ and $\mathcal{O}(10^{-2})$ for the scale-independent and peaked Bogoliubov coefficients, respectively. For the DECIGO/BBO case, assuming the scale-invariant power spectrum, we have $\bar\Omega_{\rm GW}=\mathcal{O}(10^{-16})$ for $r=10^{-2}$ and hence $C_\ell^{\rm GW}/N_\ell^{\rm GW}\leq\mathcal{O}(10^2)$.
Therefore, if DECIGO/BBO would detect SGWBs, it could be possible to test the vacuum state during inflation by the anisotropies of those SGWBs.
Note that the maximum possible value of the SNR depends on the cutoff scale, as we have discussed.
For the PTA case, assuming that the amplitude of the SGWBs is that of reported signals with recent PTA experiments, i.e. $\bar\Omega_{\rm GW}=\mathcal{O}(10^{-9})$ at nHz, we can expect $C_\ell^{\rm GW}/N_\ell^{\rm GW}\leq O(10^6)$, and 
hence, if the reported nHz SGWB is sourced by the inflationary GWs with non-Bunch-Davies states, in future PTA experiments such as SKA it could be possible to test the vacuum state by the SGWB anisotropies, similarly to DECIGO/BBO.

In the following, let us briefly investigate the detectability by making use of the cross-correlation between GWs and CMB. The SNR for the cross-correlation can be roughly computed as~\cite{Dimastrogiovanni:2021mfs}:
\begin{align}
{\rm SNR}\simeq\frac{C^{\rm GW-T}_\ell}{\sqrt{(C^{\rm GW}_\ell+N^{\rm GW}_\ell) C^{\rm TT}_\ell}},
\end{align}
where $C_\ell^{\rm TT}$
is the angular power spectrum of the CMB temperature fluctuation:
\begin{align}
C^{\rm TT}_\ell=\frac{1}{2\ell+1}\sum_m \langle a^{T}_{\ell m}a^{T*}_{\ell m}\rangle\simeq\frac{2\pi}{25\ell(\ell+1)}\mathcal{P}_\zeta \simeq 2 \times 10^{-11} \left[ \frac{30}{\ell (\ell + 1)}\right] \left( \frac{\mathcal{P}_\zeta}{2 \times 10^{-9}} \right).
\end{align}
Here, we have used the approximated form of the transfer function of the temperature fluctuation in Eq.~(\ref{eq: approx-transfer}). We then find that the SNR can be $\mathcal{O}(1)$ for $f^{ss}_{{\rm NL},0}\geq\mathcal{O}(10^6)$. For the SKA case (i.e. $N_\ell=\mathcal{O}(10^{-26})$), by assuming $\bar\Omega_{\rm GW}=\mathcal{O}(10^{-9})$, the above SNR can be $\mathcal{O}(1)$ for $f^{ss}_{{\rm NL},0}\geq\mathcal{O}(1)$.  Thus, it could be possible to test the vacuum state during inflation by the cross-correlation as well as the auto-correlation.
Note that, similarly to the SKA case, one can consider the tensor spectrum peaked at the frequency band targeted by DECIGO/BBO to enhance $\bar\Omega_{\rm GW}$ (i.e. suppress $N^{\rm GW}_\ell$) without contradicting CMB experiments by employing the scale-dependent Bogoliubov coefficient.
For such a case, we need to estimate the noise spectrum with the peaked tensor power spectrum for the DECIGO/BBO experiments
and also investigate the backreaction problem carefully. We will leave the detailed analysis of these points to our future work. 

Before closing this section, let us mention the detectability by DECIGO/BBO for higher cutoff scales.
In the present paper, we chose the lowest cutoff scale to obtain the most conservative constraint on the excitation time, but the backreaction constraint, the constraint on the non-linearity parameter, and the maximum value of the SNR are all sensitive to the choice of the cutoff scale. 
Indeed, the maximum possible value of the SNR for a higher cutoff scale can be smaller than unity, and thus it is important to investigate a strong coupling scale for each inflation model to discuss the detectability of the SGWB anisotropies predicted from each one.

\section{Summary}\label{Sec: conclusion}

The SGWB anisotropies have recently attracted a lot of attention as a tool to distinguish the source of SGWBs and a probe of the new physics in the early universe.
In this paper, we have investigated the SGWB anisotropies induced by the primordial non-Gaussianity during inflation.
Although it has been mentioned that the characteristic primordial non-Gaussianity with non-Bunch-Davies states, which brings the coupling between short- and long-wavelength modes, can induce the spatial anisotropies on SGWB,
precise theoretical predictions were not made in previous studies.
In this paper, we have explicitly formulated the primordial STT bispectrum induced during inflation in the context of Horndeski theory with non-Bunch-Davies states and investigated the STT bispectrum in the squeezed limit, which characterizes the coupling between long-wavelength scalar modes and short-wavelength tensor modes.
In the formulation, 
we carefully clarified the Trans-Planckian and strong coupling problems for the short-wavelength tensor modes in the non-Bunch-Davies state and then derived the condition to avoid them. In particular, those problems were found to be crucial for the modes satisfying $k_S/k_L>\mathcal{O}(10)$. For the SGWB anisotropies,
a large hierarchy between long- and short-wavelength modes
(roughly estimated as $k_{\rm S}/k_{\rm L} = O(10^8)$ (for PTA experiments) and $O(10^{16})$ (for DECIGO/BBO experiments)) is required, and thus close attention must be paid to them. Here, even in the CMB observations with the relatively small hierarchy between the longest and shortest modes in the observational range (roughly estimated as $k_{\rm S}/k_{\rm L}=\mathcal{O}(10^3)$), such issues are crucial. Therefore, depending on the theory and the model that determine the strong coupling scale, one must pay attention to them.
We then found that by avoiding both the Trans-Planckian and strong coupling problems the strong scale-dependent enhancement of the STT bispectrum in the squeezed limit in the case with the non-Bunch-Davies states, which was considered in the previous works, cannot be expected.
Instead, another scale dependence appears in the STT bispectrum which also depends on the non-Bunch-Davies excitation time.

To obtain the theoretical predictions for the SGWB anisotropies,
with future GW observations in mind, we assumed the two configurations of the Bogoliubov coefficients: one for the highly blue-tilted tensor spectrum, and the other for the almost scale-invariant spectrum. Using those configurations, we computed the self-correlation function of the SGWB anisotropies and the cross-correlation function of those with the CMB temperature fluctuations. Then, we evaluated the maximum possible values of the amplitudes of those correlation functions in light of the backreaction constraint and the perturbativity condition. As a result, we have found that both anisotropies can dominate those originating from the late-time metric perturbations, so-called SW effects, and shown that the anisotropies with the maximum amplitudes can potentially be tested with the DECIGO/BBO and PTA experiments. Here, ${\rm SNR}\geq1$ for DECIGO/BBO requires a
relatively large non-linearity parameter such as $\mathcal{O}(10^{6})$, compared to the PTA experiments case which requires the non-linearity parameter to be $\geq\mathcal{O}(1)$ for ${\rm SNR}\geq1$. Such a large value of the non-linearity parameter might imply the necessity of taking into account higher-point correlation functions such as a trispectrum in evaluating the non-Gaussianities. Since how large the SNR can be depends on the magnitude of $\bar\Omega_{\rm GW}$, the anisotropies with sufficiently large amplitudes detected by DECIGO/BBO would be obtained with a smaller value of the non-linearity parameter if a relatively large tensor power spectrum compared to the scale-invariant case is obtained. Therefore, it would be worth applying the peaked Bogoliubov coefficient to SGWBs at deci-Hz. Also, our analysis of the detectability is based on order estimations, and hence it would be important to compute the noise spectrum explicitly and discuss the detectability in detail. In addition, it would be interesting to discuss the ability to distinguish between models with future experiments based on the SGWB anisotropies. We will leave those to our future work.

\section*{Acknowledgements}
We thank Chunshan Lin and Giorgio Orlando for fruitful discussions. We thank the anonymous referee for useful comments. This work is supported by the grant No.~UMO-2021/42/E/ST9/00260 from the National Science Centre, Poland (S.A.), MEXT-JSPS Grant-in-Aid for Transformative Research Areas (A) ``Extreme Universe’', No.~JP21H05189 (S.A), and JSPS KAKENHI Grants No.~JP20K03968 (S.Y.), JP23H00108 (S.Y.) and JP24K00627 (S.Y.).

\appendix

\section{Perturbed actions}\label{App: perturbed-action}
\subsection{Quadratic actions}
By substituting the perturbed metric Eq.~(\ref{eq: perturbed-metric}) into the action Eq.~(\ref{eq: Horndeski action}) and expanding up to the quadratic order, the quadratic action has been obtained as~\cite{Kobayashi:2011nu}:
\begin{align}
S^{(2)}&=\int\D t\D^3x a^3\biggl[-3\mathcal{G}_T\dot\zeta^2+\frac{\mathcal{F}_T}{a^2}(\partial_i\zeta)^2+\Sigma\alpha^2-2\Theta\alpha\frac{\partial^2\beta}{a^2}+2\mathcal{G}_T\dot\zeta\frac{\partial^2\beta}{a^2}+6\Theta\alpha\dot\zeta-2\mathcal{G}_T\alpha\frac{\partial^2\zeta}{a^2}\notag\\
&\quad\quad\quad\quad\quad\quad\ +\frac{\mathcal{G}_T}{8}\dot h_{ij}^2-\frac{\mathcal{F}_T}{a^2}(\partial_k h_{ij})^2\biggr],
\end{align}
\begin{align}
\Theta&:=-\dot{\phi}XG_{3X}+2HG_4-8HXG_{4X}\notag-8HX^2G_{4XX}+\dot{\phi}G_{4\phi}+2X\dot{\phi}G_{4\phi{X}}\notag\\&
\quad-H^2\dot{\phi}(5XG_{5X}+2X^2G_{5XX})+2HX(3G_{5\phi}+2XG_{5\phi{X}}), \label{Theta-ap}\\
\Sigma&:=XG_{2X}+2X^2G_{2XX}+12H\dot{\phi}XG_{3X}+6H\dot{\phi}X^2G_{3XX}-2XG_{3\phi}-2X^2G_{3\phi{X}}-6H^2G_4\notag\\ &
\quad +6\bigl[H^2(7XG_{4X}+16X^2G_{4XX}+4X^3G_{4XXX})-H\dot{\phi}(G_{4\phi}+5XG_{4\phi{X}}+2X^2G_{4\phi{X}X})\bigr]\notag\\ &
\quad +2H^3\dot{\phi}\left(15XG_{5X}+13X^2G_{5XX}+2X^3G_{5XXX}\right)-6H^2X(6G_{5\phi}+9XG_{5\phi{X}}+2X^2G_{5\phi{X}X}). \label{Sigma-ap}
\end{align}
By varying the quadratic action with respect to $\alpha$ and $\beta$, one obtains the solutions of the constraint equations as
\begin{align}
\alpha&=\frac{\mathcal{G}_T}{\Theta}\dot\zeta,\\
\beta&=\frac{1}{a\mathcal{G}_T}\biggl(a^3\mathcal{G}_S\psi-\frac{a\mathcal{G}_T^2}{\Theta}\zeta\biggr),
\end{align}
where $\psi:=\partial^{-2}\dot\zeta$. Eliminating $\alpha$ and $\beta$ in the quadratic actions by using the solutions of the constraint equations, one obtains the quadratic actions in Eq.~(\ref{eq: quad-scalar}) and Eq.~(\ref{eq: quad-tensor}) where the coefficients are of the form,
\begin{align}
\mathcal{G}_T&:=2\left[G_4-2XG_{4X}-X(H\dot\phi G_{5X}-G_{5\phi})\right],\\
\mathcal{F}_T&:=2\left[G_4-X(\ddot\phi G_{5X}+G_{5\phi})\right],\\
\mathcal{G}_S&:=\mathcal{G}_T\biggl(\frac{\Sigma\mathcal{G}_T}{\Theta^2}+3\biggr),\\
\mathcal{F}_S&:=\frac{1}{a}\frac{\D}{\D t}\biggl(\frac{a\mathcal{G}_T^2}{\Theta}\biggr)-\mathcal{F}_T.
\end{align}

\subsection{Cubic action from STT interactions}
Similarly to the quadratic actions, after expanding the Horndeski action with respect to metric perturbations up to the cubic order, the STT-type cubic action can be obtained as~\cite{Gao:2012ib}:
\begin{align}
S_{shh}&=\int\D t\D^3x a^3\biggl[\frac{3\mathcal{G}_T}{8}\zeta\dot h_{ij}^2-\frac{\mathcal{F}_T}{8a^2}\zeta(\partial_k h_{ij})^2-\frac{\mu}{4}\dot\zeta\dot h_{ij}^2-\frac{\Gamma}{8}\alpha\dot h_{ij}^2-\frac{\mathcal{G}_T}{8a^2}\alpha(\partial_k h_{ij})^2\notag\\
&\quad\quad\quad\quad\quad\quad\ -\frac{\mu}{2a^2}\alpha\dot h_{ij}\partial^2 h_{ij}-\frac{\mathcal{G}_T}{4a^2}\partial_k\beta\dot h_{ij}\partial_k h_{ij}-\frac{\mu}{2a^2}\biggl(\partial_i\partial_j\beta\dot h_{ik}\dot h_{jk}-\frac{1}{2}\partial^2\beta\dot h_{ij}^2\biggr)\biggr],
\end{align}
where
\begin{align}
\Gamma&:=2G_4-8XG_{4X}-8X^2G_{4XX}-2H\dot\phi(5XG_{5X}+2X^2G_{5XX})+2X(3G_{5\phi}+2XG_{5\phi X}), \label{Gamma-ap}\\
&\quad-12HX(6G_{5\phi}+9XG_{5\phi X}+2X^2G_{5\phi XX}), \label{Xi-ap}\\
\mu&:=\dot\phi X G_{5X}. \label{eq: def-mu}
\end{align}
After eliminating $\alpha$ and $\beta$, one obtains
\begin{align}
S^{(3)}_{shh}=\int\D t\D^3x \biggl[a^3\mathcal{L}_{shh}+E_{shh}\biggr],
\end{align}
where $\mathcal{L}_{shh}$ is given in Eq.~(\ref{eq: cubic-lagrangian}), and $E_{shh}$ is defined by
\begin{align}
E_{shh}=\frac{\mu}{4\mathcal{G}_S}\frac{\mathcal{G}_T^2}{\Theta\mathcal{F}_T}\dot h_{ij}^2E^s+\frac{\mathcal{G}_T^2}{2\Theta\mathcal{F}_T}\biggl(\frac{\zeta}{2}+\frac{\mu}{\mathcal{G}_T}\dot\zeta\biggr)\dot h_{ij}E^h_{ij}.
\end{align}
Each coefficient of the terms in $\mathcal{L}_{shh}$ takes the following form:
\begin{align}
b_1&= \frac{3\mathcal{G}_T}{8}\biggl\{1-\frac{H\mathcal{G}_T^2}{\Theta\mathcal{F}_T}+\frac{\mathcal{G}_T}{3}\biggl[-\frac{\dot{\mathcal{F}}_T}{\mathcal{F}_T^2}\frac{\mathcal{G}_T}{\Theta}+\frac{1}{\mathcal{F}_T}\frac{\D}{\D t}\biggl(\frac{\mathcal{G}_T}{\Theta}\biggr)\biggr]\biggr\},\\
b_2&=\frac{\mathcal{F}_S}{8},\\
b_3&=-\frac{\mathcal{G}_S}{4},\\
b_4&=\frac{\mathcal{G}_T}{8\Theta\mathcal{F}_T}(\mathcal{G}_T^2-\Gamma\mathcal{F}_T)+\frac{\mu}{4}\biggl[\frac{\mathcal{G}_S}{\mathcal{G}_T}-1-\frac{H\mathcal{G}_T^2}{\Theta\mathcal{F}_T}(6+g_S)\biggr]+\frac{\mathcal{G}_T^2}{4}\frac{\D}{\D t}\biggl(\frac{\mu}{\Theta\mathcal{F}_T}\biggr),\\
b_5&=\frac{\mu\mathcal{G}_T}{4\Theta}\biggl(\frac{\mathcal{F}_S\mathcal{G}_T}{\mathcal{F}_T\mathcal{G}_S}-1\biggr),\\
b_6&=-\frac{\mu\mathcal{G}_S}{2\mathcal{G}_T},\\
b_7&=\frac{\mu\mathcal{G}_T}{2\Theta}.
\end{align}
The terms proportional to $E^s$ and $E^h_{ij}$ can be eliminated via the following field definition:
\begin{align}
h_{ij}&\to h_{ij}+\frac{\mathcal{G}_T^2}{\Theta\mathcal{F}_T}\biggl(\zeta+\frac{2\mu}{\mathcal{G}_T}\dot\zeta\biggr)\dot h_{ij},\\
\zeta&\to \zeta+\frac{\mu}{8\mathcal{G}_S}\frac{\mathcal{G}_T^2}{\Theta\mathcal{F}_T}\dot h_{ij}^2.
\end{align}

\section{Primordial STT bispectrum}\label{App: cross-bispectrum}
In this section, we summarize the details of the computations of the primordial STT bispectrum. To obtain a general expression in the context of the non-Bunch-Davies states, we assume that both scalar and tensor modes are in non-Bunch-Davies states. In the main text, we chose $\eta_0$ in such a way that the primordial GWs at the small scales avoid the Trans-Planckian problem. However, the results obtained under Eq.~(\ref{eq:ineq-TransPlanckian}) are valid at least for the primordial non-Gaussianities on CMB scales. Thus, to obtain a general expression that can be used widely, we consider possible values of $\eta_0$. We first assume that (i) the short modes of the tensor perturbations were initially in excited states, i.e. $\eta_0$ denotes the time when both long and short modes are on subhorizon scales, $-k_i\eta_0\gg1$ (i.e. $-K'\eta_0\gg1, -K_i'\eta_0\gg1$ under the squeezed limit). Then, we consider the case in which (ii) the perturbations got excited from the Bunch-Davies state when the long mode crossed the sound horizon, $-c_sk_1\eta_0=1$ (i.e. $-K'\eta_0\gg1, -K_2'\eta_0\simeq-K_3'\eta_0=1$ under the squeezed limit). We 
 also consider the case where (iii) the long mode was already on superhorizon scales, $-k_L\eta_0\ll1$, while the short modes were still on subhorizon scales, $-k_S\eta_0\gg1$, when the perturbations got excited (i.e. $-K'\eta_0\gg1, -K_2'\eta_0\simeq-K_3'\eta_0\ll1$ under the squeezed limit). Last, we assume that (iv) both long and short modes were on superhorizon scales, $-k_i\eta_0\ll1$, when the perturbations got excited (i.e. $-K'\eta_0\ll1, -K_i'\eta_0\ll1$ under the squeezed limit).

Let us parametrize the STT bispectrum as
\begin{align}
\mathcal{B}_{shh}^{s_2s_3}&={\rm Re}\Biggl[\sum^{7}_{i=1}(\alpha_{k_1}-\beta_{k_1})\left(\alpha^{(s_2)}_{k_2}-\beta^{(s_2)}_{k_2}\right)\left(\alpha^{(s_3)}_{k_3}-\beta^{(s_3)}_{k_3}\right)\frac{2}{k_1^3k_2^3k_3^3}\frac{H^6}{\mathcal{F}_S\mathcal{F}_T^2c_sc_h^2}b_i\mathcal{F}_{b_i}\mathcal{V}_{b_i}\notag\\
&\quad\quad\quad\quad\ +(s_2, k_2\leftrightarrow s_3, k_3)\Biggr].
\end{align}

\subsection{$b_1$ term}
By using the in-in formalism, the results for the $b_1$ term can be written as
\begin{align}
\mathcal{F}_{b_1}&=\frac{c_h^4k_2^2k_3^2}{H^2}\biggl[\mathcal{I}_{b_1,K'}\alpha^*_{k_1}\alpha^{(s_2)*}_{k_2}\alpha^{(s_3)*}_{k_3}+\mathcal{I}^*_{b_1,K'}\beta^*_{k_1}\beta^{(s_2)*}_{k_2}\beta^{(s_3)*}_{k_3}+\mathcal{I}_{b_1,K_1'}\beta^*_{k_1}\alpha^{(s_2)*}_{k_2}\alpha^{(s_3)*}_{k_3}\notag\\
&\quad\quad\quad\quad\quad +\mathcal{I}^*_{b_1,K_1'}\alpha^*_{k_1}\beta^{(s_2)*}_{k_2}\beta^{(s_3)*}_{k_3}+\mathcal{I}_{b_1,K_2'}\alpha^*_{k_1}\beta^{(s_2)*}_{k_2}\alpha^{(s_3)*}_{k_3}+\mathcal{I}^*_{b_1,K_2'}\beta^*_{k_1}\alpha^{(s_2)*}_{k_2}\beta^{(s_3)*}_{k_3}\notag\\
&\quad\quad\quad\quad\quad+\mathcal{I}_{b_1,K_3'}\alpha^*_{k_1}\alpha^{(s_2)*}_{k_2}\beta^{(s_3)*}_{k_3}+\mathcal{I}^*_{b_1,K_3'}\beta^*_{k_1}\beta^{(s_2)*}_{k_2}\alpha^{(s_3)*}_{k_3}\biggr],
\end{align}
and
\begin{align}
\mathcal{V}_{b_1}=\frac{1}{16k_2^2k_3^2}[k_1^2-(s_2k_2+s_3k_3)^2]^2,
\end{align}
where
\begin{align}
\mathcal{I}_{b_1,K'}&:=\int\D\eta(i+c_sk_1\eta)e^{iK'\eta},\\
\mathcal{I}_{b_1,K_1'}&:=-\int\D\eta(i-c_sk_1\eta)e^{iK_1'\eta},\\
\mathcal{I}_{b_1,K_2'}&:=-\int\D\eta(i+c_sk_1\eta)e^{iK_2'\eta},\\
\mathcal{I}_{b_1,K_3'}&:=-\int\D\eta(i+c_sk_1\eta)e^{iK_3'\eta}.
\end{align}
Here, only $\mathcal{I}_{b_1,K'}$ appears in the case of the Bunch-Davies state, and thus one can evaluate the corrections to the bispectrum due to the excited modes by computing $\mathcal{I}_{b_1,K'}/\mathcal{I}_{b_1,K_i'}$.

In case (i), we obtain
\begin{align}
\mathcal{I}_{b_1,K'}&=\mathcal{J}_{b_1}(k_1,k_2,k_3), \label{eq: Ikb1i}\\
\mathcal{I}_{b_1,K_1'}&=-\mathcal{J}_{b_1}(-k_1,k_2,k_3), \label{eq: Ik1b1i}\\
\mathcal{I}_{b_1,K_2'}&=-\mathcal{J}_{b_1}(k_1,-k_2,k_3), \label{eq: Ik2b1i}\\
\mathcal{I}_{b_1,K_3'}&=-\mathcal{J}_{b_1}(k_1,k_2,-k_3) \label{eq: Ik3b1i},
\end{align}
where
\begin{align}
\mathcal{J}_{b_1}(k_1,k_2,k_3):=\frac{c_sk_1+K'}{K'^2}.
\end{align}

In cases (ii), (iii), and (iv), $\mathcal{I}_{b_1,K'}$ and $\mathcal{I}_{b_1,K_1'}$ take the same form as in Eq.~\eqref{eq: Ikb1i} and Eq.~\eqref{eq: Ik1b1i}, respectively, and we can write the results of the other two integrals as
\begin{align}
\mathcal{I}_{b_1,K_2'}&=-\mathcal{J}_{b_1}(k_1,-k_2,k_3)+\mathcal{S}_{b_1}(k_1,-k_2,k_3),\\
\mathcal{I}_{b_1,K_3'}&=-\mathcal{J}_{b_1}(k_1,k_2,-k_3)+\mathcal{S}_{b_1}(k_1,k_2,-k_3),
\end{align}
where
\begin{align}
\mathcal{S}_{b_1}(k_1,k_2,k_3):=\frac{e^{iK'\eta_0}}{K'^2}(c_sk_1+K'-ic_sk_1K'\eta_0).
\end{align}
In case (ii), $\eta_0$ is replaced with $-1/(c_sk_1)$.

\subsection{$b_2$ term}
The results for the $b_2$ term can be written as
\begin{align}
\mathcal{F}_{b_2}&=\frac{1}{H^2}\biggl[\mathcal{I}_{b_2,K'}\alpha^*_{k_1}\alpha^{(s_2)*}_{k_2}\alpha^{(s_3)*}_{k_3}+\mathcal{I}^*_{b_2,K'}\beta^*_{k_1}\beta^{(s_2)*}_{k_2}\beta^{(s_3)*}_{k_3}+\mathcal{I}_{b_2,K_1'}\beta^*_{k_1}\alpha^{(s_2)*}_{k_2}\alpha^{(s_3)*}_{k_3}\notag\\
&\quad\quad\quad\ +\mathcal{I}^*_{b_2,K_1'}\alpha^*_{k_1}\beta^{(s_2)*}_{k_2}\beta^{(s_3)*}_{k_3}+\mathcal{I}_{b_2,K_2'}\alpha^*_{k_1}\beta^{(s_2)*}_{k_2}\alpha^{(s_3)*}_{k_3}+\mathcal{I}^*_{b_2,K_2'}\beta^*_{k_1}\alpha^{(s_2)*}_{k_2}\beta^{(s_3)*}_{k_3}\notag\\
&\quad\quad\quad\ +\mathcal{I}_{b_2,K_3'}\alpha^*_{k_1}\alpha^{(s_2)*}_{k_2}\beta^{(s_3)*}_{k_3}+\mathcal{I}^*_{b_2,K_3'}\beta^*_{k_1}\beta^{(s_2)*}_{k_2}\alpha^{(s_3)*}_{k_3}\biggr],
\end{align}
and
\begin{align}
\mathcal{V}_{b_2}=\frac{k_1^2-k_2^2-k_3^2}{2}\mathcal{V}_{b_1},
\end{align}
where
\begin{align}
\mathcal{I}_{b_2,K'}&:=\int\D\eta\frac{1}{\eta^2}(i+c_sk_1\eta)(i+c_hk_2\eta)(i+c_hk_3\eta)e^{iK'\eta},\\
\mathcal{I}_{b_2,K_1'}&:=\int\D\eta\frac{1}{\eta^2}(-i+c_sk_1\eta)(i+c_hk_2\eta)(i+c_hk_3\eta)e^{iK_1'\eta},\\
\mathcal{I}_{b_2,K_2'}&:=\int\D\eta\frac{1}{\eta^2}(i+c_sk_1\eta)(-i+c_hk_2\eta)(i+c_hk_3\eta)e^{iK_2'\eta},\\
\mathcal{I}_{b_2,K_3'}&:=\int\D\eta\frac{1}{\eta^2}(i+c_sk_1\eta)(i+c_hk_2\eta)(-i+c_hk_3\eta)e^{iK_3'\eta},
\end{align}

In case (i), we obtain
\begin{align}
\mathcal{I}_{b_2,K'}&=\mathcal{J}_{b_2}(k_1,k_2,k_3), \label{eq: Ikb2}\\
\mathcal{I}_{b_2,K_1'}&=-\mathcal{J}_{b_2}(-k_1,k_2,k_3), \label{eq: Ik1b2}\\
\mathcal{I}_{b_2,K_2'}&=-\mathcal{J}_{b_2}(k_1,-k_2,k_3),\\
\mathcal{I}_{b_2,K_3'}&=-\mathcal{J}_{b_2}(k_1,k_2,-k_3),
\end{align}
where
\begin{align}
\mathcal{J}_{b_2}(k_1,k_2,k_3)&:=-\frac{1}{K'^2}[c_s^3k_1^3+2c_s^2c_hk_1^2(k_2+k_3)+2c_sc_h^2k_1(k_2^2+k_2k_3+k_3^2)\notag\\
&\quad\quad\quad\quad\ +c_h^3(k_2+k_3)(k_2^2+k_2k_3+k_3^2)],
\end{align}
and we used the fact that the term proportional to $1/\eta_*$, with $\eta_*(\to0)$ denoting the conformal time at the end of inflation, does not contribute to the three-point function because of the following equation:
\begin{align}
&{\rm Re}\biggl[i(\alpha_{k_1}-\beta_{k_1})\left(\alpha^{(s_2)}_{k_2}-\beta^{(s_2)}_{k_2}\right)\left(\alpha^{(s_3)}_{k_3}-\beta^{(s_3)}_{k_3}\right)(\alpha^*_{k_1}-\beta^*_{k_1})\left(\alpha^{(s_2)*}_{k_2}-\beta^{(s_2)*}_{k_2}\right)\notag\\
&\quad\ \times\left(\alpha^{(s_3)*}_{k_3}-\beta^{(s_3)*}_{k_3}\right)\biggr]=0.
\end{align}

In the other three cases, $\mathcal{I}_{b_2,K'}$ and $\mathcal{I}_{b_2,K_1'}$ take the same form as in Eq.~\eqref{eq: Ikb2} and Eq.~\eqref{eq: Ik1b2}, respectively, and we can write the results of the other two integrals as
\begin{align}
\mathcal{I}_{b_2,K_2'}&=-\mathcal{J}_{b_2}(k_1,-k_2,k_3)+\mathcal{S}_{b_2}(k_1,-k_2,k_3),\\
\mathcal{I}_{b_2,K_3'}&=-\mathcal{J}_{b_2}(k_1,k_2,-k_3)+\mathcal{S}_{b_2}(k_1,k_2,-k_3),
\end{align}
where
\begin{align}
\mathcal{S}_{b_2}(k_1,k_2,k_3)&:=e^{iK'\eta_0}\biggl[\frac{c_h}{K'^2}(c_h^2k_2k_3(k_2+k_3)+c_s^2k_1^2(k_2+k_3)+c_sc_hk_1(k_2^2+4k_2k_3+k_3^2))\notag\\
&\quad\quad\quad\quad\ +\frac{i}{\eta_0}-\frac{ic_sc_h^2k_1k_2k_3\eta_0}{K'}\biggr].
\end{align}
In case (ii), $\eta_0$ is replaced with $-1/(c_sk_1)$.

\subsection{$b_3$ term}
The results for the $b_3$ term can be written as
\begin{align}
\mathcal{F}_{b_3}&=\frac{c_h^2c_s^2k_1^2k_2^2}{H^2}\biggl[\mathcal{I}_{b_3,K'}\alpha^*_{k_1}\alpha^{(s_2)*}_{k_2}\alpha^{(s_3)*}_{k_3}+\mathcal{I}^*_{b_3,K'}\beta^*_{k_1}\beta^{(s_2)*}_{k_2}\beta^{(s_3)*}_{k_3}+\mathcal{I}_{b_3,K_1'}\beta^*_{k_1}\alpha^{(s_2)*}_{k_2}\alpha^{(s_3)*}_{k_3}\notag\\
&\quad\quad\quad\quad\quad\ +\mathcal{I}^*_{b_3,K_1'}\alpha^*_{k_1}\beta^{(s_2)*}_{k_2}\beta^{(s_3)*}_{k_3}+\mathcal{I}_{b_3,K_2'}\alpha^*_{k_1}\beta^{(s_2)*}_{k_2}\alpha^{(s_3)*}_{k_3}+\mathcal{I}^*_{b_3,K_2'}\beta^*_{k_1}\alpha^{(s_2)*}_{k_2}\beta^{(s_3)*}_{k_3}\notag\\
&\quad\quad\quad\quad\quad\ +\mathcal{I}_{b_3,K_3'}\alpha^*_{k_1}\alpha^{(s_2)*}_{k_2}\beta^{(s_3)*}_{k_3}+\mathcal{I}^*_{b_3,K_3'}\beta^*_{k_1}\beta^{(s_2)*}_{k_2}\alpha^{(s_3)*}_{k_3}\biggr],
\end{align}
and
\begin{align}
\mathcal{V}_{b_3}=-\frac{k_1^2-k_2^2+k_3^2}{2k_1^2}\mathcal{V}_{b_1},
\end{align}
where
\begin{align}
\mathcal{I}_{b_3,K'}&:=\int\D\eta (i+c_hk_3\eta)e^{iK'\eta},\\
\mathcal{I}_{b_3,K_1'}&:=-\int\D\eta (i+c_hk_3\eta)e^{iK_1'\eta},\\
\mathcal{I}_{b_3,K_2'}&:=-\int\D\eta (i+c_hk_3\eta)e^{iK_2'\eta},\\
\mathcal{I}_{b_3,K_3'}&:=-\int\D\eta (i-c_hk_3\eta)e^{iK_3'\eta}.
\end{align}
In case (i), we obtain
\begin{align}
\mathcal{I}_{b_3,K'}&:=\mathcal{J}_{b_3}(k_1,k_2,k_3), \label{eq: Ikb3}\\
\mathcal{I}_{b_3,K_1'}&:=-\mathcal{J}_{b_3}(-k_1,k_2,k_3), \label{eq: Ik1b3}\\
\mathcal{I}_{b_3,K_2'}&:=-\mathcal{J}_{b_3}(k_1,-k_2,k_3),\\
\mathcal{I}_{b_3,K_3'}&:=-\mathcal{J}_{b_3}(k_1,k_2,-k_3),
\end{align}
where
\begin{align}
\mathcal{J}_{b_3}(k_1,k_2,k_3):=\frac{K'+c_hk_3}{K'^2}.
\end{align}

In the other three cases, $\mathcal{I}_{b_3,K'}$ and $\mathcal{I}_{b_3,K_1'}$ take the same form as in Eq.~\eqref{eq: Ikb3} and Eq.~\eqref{eq: Ik1b3}, respectively, and we can write the results of the other two integrals as
\begin{align}
\mathcal{I}_{b_3,K_2'}&:=-\mathcal{J}_{b_3}(k_1,-k_2,k_3)+\mathcal{S}_{b_3}(k_1,-k_2,k_3),\\
\mathcal{I}_{b_3,K_3'}&:=-\mathcal{J}_{b_3}(k_1,k_2,-k_3)+\mathcal{S}_{b_3}(k_1,k_2,-k_3),
\end{align}
where
\begin{align}
\mathcal{S}_{b_3}(k_1,k_2,k_3):=\frac{e^{iK'\eta_0}}{K'^2}(K'+c_hk_3-ic_hK'k_3\eta_0).
\end{align}
In case (ii), $\eta_0$ is replaced with $-1/(c_sk_1)$.

\subsection{$b_4$ term}
The results for the $b_4$ term can be written as
\begin{align}
\mathcal{F}_{b_4}&=\frac{c_h^4c_s^2k_1^2k_2^2k_3^2}{H}\biggl[\mathcal{I}_{b_4,K'}\alpha^*_{k_1}\alpha^{(s_2)*}_{k_2}\alpha^{(s_3)*}_{k_3}+\mathcal{I}^*_{b_4,K'}\beta^*_{k_1}\beta^{(s_2)*}_{k_2}\beta^{(s_3)*}_{k_3}+\mathcal{I}_{b_4,K_1'}\beta^*_{k_1}\alpha^{(s_2)*}_{k_2}\alpha^{(s_3)*}_{k_3}\notag\\
&\quad\quad\quad\quad\quad\quad\ +\mathcal{I}^*_{b_4,K_1'}\alpha^*_{k_1}\beta^{(s_2)*}_{k_2}\beta^{(s_3)*}_{k_3}+\mathcal{I}_{b_4,K_2'}\alpha^*_{k_1}\beta^{(s_2)*}_{k_2}\alpha^{(s_3)*}_{k_3}+\mathcal{I}^*_{b_4,K_2'}\beta^*_{k_1}\alpha^{(s_2)*}_{k_2}\beta^{(s_3)*}_{k_3}\notag\\
&\quad\quad\quad\quad\quad\quad\ +\mathcal{I}_{b_4,K_3'}\alpha^*_{k_1}\alpha^{(s_2)*}_{k_2}\beta^{(s_3)*}_{k_3}+\mathcal{I}^*_{b_4,K_3'}\beta^*_{k_1}\beta^{(s_2)*}_{k_2}\alpha^{(s_3)*}_{k_3}\biggr],
\end{align}
and
\begin{align}
\mathcal{V}_{b_4}=\mathcal{V}_{b_1},
\end{align}
where
\begin{align}
\mathcal{I}_{b_4,K'}&:=-i\int\D\eta\eta^2e^{iK'\eta},\\
\mathcal{I}_{b_4,K_1'}&:=i\int\D\eta\eta^2e^{iK_1'\eta},\\
\mathcal{I}_{b_4,K_2'}&:=i\int\D\eta\eta^2e^{iK_2'\eta},\\
\mathcal{I}_{b_4,K_3'}&:=i\int\D\eta\eta^2e^{iK_3'\eta}.
\end{align}
In case (i), we have
\begin{align}
\mathcal{I}_{b_4,K'}&=\mathcal{J}_{b_4}(k_1,k_2,k_3), \label{eq: Ikb4}\\
\mathcal{I}_{b_4,K_1'}&=-\mathcal{J}_{b_4}(-k_1,k_2,k_3), \label{eq: Ik1b4}\\
\mathcal{I}_{b_4,K_2'}&=-\mathcal{J}_{b_4}(k_1,-k_2,k_3),\\
\mathcal{I}_{b_4,K_3'}&=-\mathcal{J}_{b_4}(k_1,k_2,-k_3),
\end{align}
where
\begin{align}
\mathcal{J}_{b_4}(k_1,k_2,k_3):=\frac{2}{K'^3}.
\end{align}

In the other three cases, $\mathcal{I}_{b_4,K'}$ and $\mathcal{I}_{b_4,K_1'}$ take the same form as in Eq.~\eqref{eq: Ikb4} and Eq.~\eqref{eq: Ik1b4}, respectively, and we can write the results of the other two integrals as
\begin{align}
\mathcal{I}_{b_4,K_2'}&:=-\mathcal{J}_{b_4}(k_1,-k_2,k_3)+\mathcal{S}_{b_4}(k_1,-k_2,k_3),\\
\mathcal{I}_{b_4,K_3'}&:=-\mathcal{J}_{b_4}(k_1,k_2,-k_3)+\mathcal{S}_{b_4}(k_1,k_2,-k_3),
\end{align}
where
\begin{align}
\mathcal{S}_{b_4}(k_1,k_2,k_3):=-\frac{i}{K'^3}e^{iK'\eta_0}(2i+2K'\eta_0-iK'^2\eta_0^2)
\end{align}
In case (ii), $\eta_0$ is replaced with $-1/(c_sk_1)$.

\subsection{$b_5$ term}
The results for the $b_5$ term can be written as
\begin{align}
\mathcal{F}_{b_5}&=c_h^4k_1^2k_2^2k_3^2\biggl[\mathcal{I}_{b_5,K'}\alpha^*_{k_1}\alpha^{(s_2)*}_{k_2}\alpha^{(s_3)*}_{k_3}+\mathcal{I}^*_{b_5,K'}\beta^*_{k_1}\beta^{(s_2)*}_{k_2}\beta^{(s_3)*}_{k_3}+\mathcal{I}_{b_5,K_1'}\beta^*_{k_1}\alpha^{(s_2)*}_{k_2}\alpha^{(s_3)*}_{k_3}\notag\\
&\quad\quad\quad\quad\quad\ +\mathcal{I}^*_{b_5,K_1'}\alpha^*_{k_1}\beta^{(s_2)*}_{k_2}\beta^{(s_3)*}_{k_3} +\mathcal{I}_{b_5,K_2'}\alpha^*_{k_1}\beta^{(s_2)*}_{k_2}\alpha^{(s_3)*}_{k_3}+\mathcal{I}^*_{b_5,K_2'}\beta^*_{k_1}\alpha^{(s_2)*}_{k_2}\beta^{(s_3)*}_{k_3}\notag\\
&\quad\quad\quad\quad\quad\ +\mathcal{I}_{b_5,K_3'}\alpha^*_{k_1}\alpha^{(s_2)*}_{k_2}\beta^{(s_3)*}_{k_3}+\mathcal{I}^*_{b_5,K_3'}\beta^*_{k_1}\beta^{(s_2)*}_{k_2}\alpha^{(s_3)*}_{k_3}\biggr],
\end{align}
and
\begin{align}
\mathcal{V}_{b_5}=k_1^2\mathcal{V}_{b_1},
\end{align}
where
\begin{align}
\mathcal{I}_{b_5,K'}&:=-\int\D\eta\eta^2(i+c_sk_1\eta)e^{iK'\eta},\\
\mathcal{I}_{b_5,K_1'}&:=-\int\D\eta\eta^2(-i+c_sk_1\eta)e^{iK_1'\eta},\\
\mathcal{I}_{b_5,K_2'}&:=\int\D\eta\eta^2(i+c_sk_1\eta)e^{iK_2'\eta},\\
\mathcal{I}_{b_5,K_3'}&:=\int\D\eta\eta^2(i+c_sk_1\eta)e^{iK_3'\eta}.
\end{align}
In case (i), we have
\begin{align}
\mathcal{I}_{b_5,K'}&=\mathcal{J}_{b_5}(k_1,k_2,k_3), \label{eq: Ikb5}\\
\mathcal{I}_{b_5,K_1'}&=-\mathcal{J}_{b_5}(-k_1,k_2,k_3), \label{eq: Ik1b5}\\
\mathcal{I}_{b_5,K_2'}&=-\mathcal{J}_{b_5}(k_1,-k_2,k_3),\\
\mathcal{I}_{b_5,K_3'}&=-\mathcal{J}_{b_5}(k_1,k_2,-k_3),
\end{align}
where
\begin{align}
\mathcal{J}_{b_5}(k_1,k_2,k_3):=\frac{2(3c_sk_1+K')}{K'^4}.
\end{align}

In the other three cases, $\mathcal{I}_{b_5,K'}$ and $\mathcal{I}_{b_5,K_1'}$ take the same form as in Eq.~\eqref{eq: Ikb5} and Eq.~\eqref{eq: Ik1b5}, respectively, and we can write the results of the other two integrals as
\begin{align}
\mathcal{I}_{b_5,K_2'}&:=-\mathcal{J}_{b_5}(k_1,-k_2,k_3)+\mathcal{S}_{b_5}(k_1,-k_2,k_3),\\
\mathcal{I}_{b_5,K_3'}&:=-\mathcal{J}_{b_5}(k_1,k_2,-k_3)+\mathcal{S}_{b_5}(k_1,k_2,-k_3),
\end{align}
where
\begin{align}
\mathcal{S}_{b_5}(k_1,k_2,k_3):=e^{iK'\eta_0}\biggl\{\frac{2(3c_sk_1+K')}{K'^4}+\frac{i\eta_0}{K'^3}[K'(-2+iK'\eta_0)+c_sk_1(-6+3iK'\eta_0+K'^2\eta_0^2)]\biggr\}.
\end{align}
In case (ii), $\eta_0$ is replaced with $-1/(c_sk_1)$.

\subsection{$b_6$ term}
The results for the $b_6$ term can be written as
\begin{align}
\mathcal{F}_{b_6}=\mathcal{F}_{b_4},
\end{align}
and
\begin{align}
\mathcal{V}_{b_6}=\frac{K}{32k_1^2k_2^2k_3^2}(k_1-k_2-k_3)(k_1+k_2-k_3)(k_1-k_2+k_3)[k_1^2-(s_2k_2+s_3k_3)^2],
\end{align}
where $K=k_1+k_2+k_3$.

\subsection{$b_7$ term}
The results for the $b_7$ term can be written as
\begin{align}
\mathcal{F}_{b_7}=\mathcal{F}_{b_5},
\end{align}
and
\begin{align}
\mathcal{V}_{b_7}=k_1^2\mathcal{V}_{b_6}.
\end{align}

\subsection{Field redefinition}
Since the field redefinition does not bring any contributions from the far past to the resultant bispectrum, let us consider the superhorizon scales on which we have
\begin{align}
\dot h^{(s)}_{{\bf k}}\simeq ic_h^2k^2\eta^2Hh^{(s)}_{\bf k}.
\end{align}
The quadratic-order terms in the field redefinition involve at least one time-derivative of the tensor perturbations, which means that those do not contribute to the bispectrum evaluated at the end of inflation $\eta\to0$ (i.e. $\dot h^{(s)}_{{\bf k}}\to0$). Therefore, the field redefinition always yields subleading contributions.

\section{$\mathcal{V}_{b_i}$ at the squeezed limit}\label{App: squeezed-limit-polarization-tensor}
One can take the squeezed limit to $\mathcal{V}_{b_i}$ straightforwardly as
\begin{align}
    \mathcal{V}_{b_1} &=
\frac{1}{16} \left( s_2 + s_3 \right)^4 - \frac{1}{8} \left( s_2 +  s_3 \right)^3 \left( s_2 - s_3 \right) (\hat n\cdot\hat q)
\frac{q}{k}\notag\\
&\quad\ -\frac{1}{8} \left( s_2 + s_3 \right)^2 \left[ 1 - (\hat n\cdot\hat q)^2 \left( s_2^2 - s_2 s_3 + s_3^2 \right)   \right]
    \frac{q^2}{k^2} + O((q/k)^3),\\
\mathcal{V}_{b_2}&=
-\frac{1}{16} \left( s_2 + s_3 \right)^2 k^2 \notag \\
&\quad \times \biggl\{ \left( s_2 + s_3 \right)^2 - 2(\hat n\cdot\hat q) (s_2 - s_3)(s_2 + s_3) \frac{q}{k} -
\biggl[8+(1-8(\hat n\cdot\hat q)^2)s_2^2+(1-8 (\hat n\cdot\hat q)^2)s_3^2 \notag\\
&\quad\quad\ + 2 s_2(s_3 +4 (\hat n\cdot\hat q)^2 s_3) \frac{q^2}{4k^2}\biggr] + O((q/k)^3)\biggr\},\\
    \mathcal{V}_{b_3} 
&= - \frac{(\hat n\cdot\hat q)}{16} \frac{k}{q}
\left( s_2 + s_3 \right)^2 \notag\\
&\quad\times\left\{
\left( s_2 + s_3 \right)^2 + \Big[s_2 - 4(\hat n\cdot\hat q)^2 s_2 + s_3 + 4 (\hat n\cdot\hat q)^2 s_3\Big] (s_2 + s_3)\frac{q}{2k (\hat n\cdot\hat q)} \right. \\
&\quad\quad\ \left. - \Big[2+ (1-2(\hat n\cdot\hat q)^2)s_2^2 + 2 (\hat n\cdot\hat q)^2 s_2 s_3 - (1+2 (\hat n\cdot\hat q)^2)s_3^2\Big] \frac{q^2}{k^2}
+ O((q/k)^3)
\right\},\\
    \mathcal{V}_{b_4} &= \mathcal{V}_{b_1},\\
    \mathcal{V}_{b_5} &
= q^2 \mathcal{V}_{b_1},\\
    \mathcal{V}_{b_6} &= 
\frac{1}{8} [1 - (\hat n\cdot\hat q)^2] (s_2 + s_3)^2
- (\hat n\cdot\hat q) [1-(\hat n\cdot\hat q)^2] (s_2^2 - s_3^2) \frac{q}{8k} \\
&\quad - [1-(\hat n\cdot\hat q)^2]\Big\{4 +[1-4(\hat n\cdot\hat q)^2]s_2^2 +2[1-(\hat n\cdot\hat q)^2]s_2 s_3 + [1-4(\hat n\cdot\hat q)^2] s_3^2\Big\} \frac{q^2}{32k^2}\notag\\
&\quad + O((q/k)^3),\\
    \mathcal{V}_{b_7} &
= q^2 \mathcal{V}_{b_6}.
\end{align}
We have used the notation in Sec.~\ref{sec: general expression}. Also, $q$ and $k$ denote the wavenumbers of the long and short modes, respectively.
Based on these expressions,
one can find that in the case with $s_2 = - s_3$,
$\mathcal{V}_{b_6}$ is suppressed at most
in the order of $q/k$ and the others are more suppressed. Therefore, $\mathcal{V}_{b_i}$ take their maximum values for $s_2=s_3$.

\section{A relation between the enhancement of the SGWB anisotropies and the suppression of a scalar non-Gaussianity}\label{App: SGWBanisotropies-and-scalarnon-Gaussianity}
 In this section, we first discuss whether we can enhance the SGWB anisotropies, while keeping the scalar non-Gaussianity small such that the observational bounds are satisfied, within the cubic Galileon theory whose Lagrangian density is $\mathcal{L}=G_2(\phi,X)-G_3(\phi,X)\Box\phi+(\mpl^2/2)R$. Since we have assumed the Bunch-Davies initial condition for the curvature perturbation, we use the expression of the scalar auto-bispectrum from generalized G-inflation with the Bunch-Davies state~\cite{Gao:2011qe,DeFelice:2011zh}. To discuss the above, we introduce the non-linearity parameter of the scalar non-Gaussianity:
\begin{align}
f_{\rm NL}:=\frac{10}{3}\frac{\mathcal{A}_s}{\sum_i k_i^3},
\end{align}
where $\mathcal{A}_s$ is of the form~\cite{Gao:2011qe,DeFelice:2011zh},
\begin{align}
\mathcal{A}_s&=\biggl(\frac{3\Lambda_1}{2}+\frac{3\Lambda_4}{c_s^2}\biggr)\frac{(k_1k_2k_3)^2}{K^3}+\frac{\Lambda_2}{4K}\biggl(2\sum_{i>j}k_i^2k_j^2-\frac{1}{K}\sum_{i\neq j}k_i^2k_j^3\biggr)\notag\\
&\quad+\frac{\Lambda_3}{8c_s^2}\biggl(\sum_i k_i^3+\frac{4}{K}\sum_{i>j}k_i^2k_j^2-\frac{2}{K^2}\sum_{i\neq j}k_i^2k_j^3\biggr)+\frac{\Lambda_5}{8}\biggl(\sum_i k_i^3-\frac{1}{2}\sum_{i\neq j}k_i k_j^2-\frac{2}{K^2}\sum_{i\neq j}k_i^2k_j^3\biggr)\notag\\
&\quad +\frac{\Lambda_6}{8}\frac{1}{K^2}\biggl(2\sum_i k_i^5+\sum_{i\neq j}k_i k_j^4-3\sum_{i\neq j}k_i^2k_j^3-2k_1k_2k_3\sum_{i>j}k_1k_j\biggr)\notag\\
&\quad +\frac{3\Lambda_7}{8c_s^4}\frac{1}{K}\biggl(\sum_i k_i^4-2\sum_{i<j}k_i^2k_j^2\biggr)\biggl(1+\frac{1}{K^2}\sum_{i>j}k_ik_j+\frac{3k_1k_2k_3}{K^3}\biggr)\notag\\
&\quad +\frac{\Lambda_8}{16c_s^2}\frac{1}{K^2}\biggl(7K\sum_i k_i^4+3k_1k_2k_3\sum_i k_i^2-2\sum_i k_i^5-5k_1k_2k_3K^2-12\sum_{i\neq j}k_i^2k_j^3\biggr),
\end{align}
with $K:=k_1+k_2+k_3$ and
\begin{align}
\Lambda_1&=H\biggl[\frac{\mathcal{G}_T}{\Theta}\biggl(\frac{\mathcal{G}_S}{\mathcal{F}_S}+3\frac{\mathcal{G}_T}{\mathcal{G}_S}-1\biggr)+\frac{\Xi\mathcal{G}_T}{3\Theta^2}\biggl(3\frac{\mathcal{G}_T}{\mathcal{G}_S}-1\biggr)+2\mu\biggl(\frac{1}{\mathcal{G}_S}-\frac{1}{\mathcal{G}_T}\biggr)+\frac{\Gamma}{\Theta}\biggl(3\frac{\mathcal{G}_T}{\mathcal{G}_S}-2\biggr)\notag\\
&\quad\ +\frac{2}{3}\frac{\mathcal{G}_T^3}{\Theta^3\mathcal{G}_S}(\Sigma-X\Sigma_X)-\frac{H}{3}\frac{\mathcal{G}_T^3\Xi}{\Theta^3\mathcal{G}_S}\biggr],\\
\Lambda_2&=3-\frac{H\mathcal{G}_T\mathcal{G}_S}{\mathcal{F}_S\Theta}(3-g_T+f_S+f_\Theta),\\
\Lambda_3&=\frac{\mathcal{F}_T}{\mathcal{G}_S}+\frac{H\mathcal{G}_T}{\Theta}(1+g_T+g_S-f_\Theta)-\frac{H\mathcal{G}_T^2}{\mathcal{G}_S\Theta}(1+2g_T-f_\Theta),\\
\Lambda_4&=H^2\biggl[\frac{\Xi}{3}\frac{\mathcal{G}_T^3}{\mathcal{G}_S\Theta^3}+6\mu\frac{\mathcal{G}_T}{\mathcal{G}_S\Theta}+(3\Gamma-\mathcal{G}_T)\frac{\mathcal{G}_T^2}{\mathcal{G}_S\Theta^2}\biggr],\\
\Lambda_5&=-\frac{1}{2}\frac{\mathcal{G}_S}{\mathcal{G}_T}-\frac{H}{2}\frac{\Gamma\mathcal{G}_S}{\mathcal{G}_T\Theta}(3+g_T-f_\Gamma+f_\Theta)-\mu H\frac{\mathcal{G}_S}{\mathcal{G}_T^2}(3+2g_T-f_\mu),\\
\Lambda_6&=\frac{3}{4}\frac{\mathcal{G}_S}{\mathcal{G}_T}-\frac{\mathcal{G}_S}{4\mathcal{G}_T}\frac{H\Gamma}{\Theta}(3+g_T-f_\Gamma+f_\Theta)-\mu H\frac{\mathcal{G}_S}{\mathcal{G}_T^2}\biggl(\frac{3}{2}+g_T-\frac{1}{2}f_\mu\biggr),\\
\Lambda_7&=\frac{H^2}{6}\biggl[\frac{\mathcal{G}_T^3}{\mathcal{G}_S\Theta^2}-\frac{H\Gamma\mathcal{G}_T^3}{\mathcal{G}_S\Theta^3}\biggl(1-3g_T+3f_\Theta-f_\Gamma+3\frac{\Theta\mathcal{F}_S}{H\mathcal{G}_T^2}\biggr)\notag\\
&\quad\ -6\mu H\frac{\mathcal{G}_T^2}{\mathcal{G}_S\Theta^2}\biggl(1-2g_T-f_\mu+2f_\Theta+2\frac{\Theta\mathcal{F}_S}{H\mathcal{G}_T^2}\biggr)\biggr],\\
\Lambda_8&=H\biggl[-\frac{\mathcal{G}_T}{\Theta}+\frac{2\mu H}{\Theta}\biggl(2+f_\Theta-f_\mu+\frac{\Theta\mathcal{F}_S}{H\mathcal{G}_T^2}\biggr)+H\frac{\Gamma\mathcal{G}_T}{\Theta^2}\biggl(1-\frac{1}{2}g_T-\frac{1}{2}f_\Gamma+f_\Theta+\frac{\Theta\mathcal{F}_S}{H\mathcal{G}_T^2}\biggr)\biggr].
\end{align}
Here, $\Xi$ is defined by
\begin{align}
\Xi&:=12\dot\phi{X}G_{3X}+6\dot\phi{X^2}G_{3XX}-12HG_4+6\biggl[2H(7XG_{4X}+16X^2G_{4XX}+4X^3G_{4XXX})-\dot\phi(G_{4\phi}\notag\\
&\quad+5XG_{4\phi{X}}+2X^2G_{4\phi{XX}})\biggr]+90H^2\dot\phi{X}G_{5X}+78H^2\dot\phi{X^2}G_{5XX}+12H^2\dot\phi{X^3}G_{5XXX}\notag\\
&\quad-12HX(6G_{5\phi}+9XG_{5\phi X}+2X^2G_{5\phi XX}).
\end{align}
In the following, we consider the cubic Galileon theory where we have $\Gamma=\mathcal{G}_T=\mathcal{F}_T=\mpl^2$ and $\mu=0$. Furthermore, lower-case letters such as $f_S$ and $f_\Theta$ are slow-roll parameters, and hence those are much smaller than unity on the slow-roll background.  The CMB experiments have put the observational bounds on $f_{\rm NL}$ at squeezed ($k_1\ll k_2\simeq k_3$), equilateral ($k_1=k_2=k_3$), and folded limits ($k_1=2k_2=2k_3$). In light of the Maldacena's consistency relation (i.e. the suppression of the scalar squeezed non-Gaussianity in the Bunch-Davies case), we consider the latter two limits to discuss whether the scalar non-Gaussianity is enhanced when we amplify the SGWB anisotropies. At these limits, we have
\begin{align}
f^{\rm eq}_{\rm NL}&=\frac{5}{81}\Lambda_1+\frac{10}{27}\Lambda_2+\frac{85}{108}\frac{\Lambda_3}{c_s^2}+\frac{10}{81}\frac{\Lambda_4}{c_s^2}-\frac{5}{27}(\Lambda_5+\Lambda_6)-\frac{65}{108}\frac{\Lambda_7}{c_s^4}-\frac{85}{216}\frac{\Lambda_8}{c_s^2},\\
f^{\rm fold}_{\rm NL}&=\frac{1}{32}\Lambda_1+\frac{23}{96}\Lambda_2+\frac{21}{32}\frac{\Lambda_3}{c_s^2}+\frac{1}{16}\frac{\Lambda_4}{c_s^2}-\frac{1}{96}\Lambda_5+\frac{1}{48}\Lambda_6.
\end{align}
Since both parameters need to be $\mathcal{O}(10)$ in light of the observational constraints obtained by Planck~\cite{Planck:2019kim}, the following combination is naively up to $\mathcal{O}(10^2)$:
\begin{align}
\tilde f_{\rm NL}&:=\frac{81}{5}f^{\rm eq}_{\rm NL}-32 f^{\rm fold}_{\rm NL}\notag\\
&=-\frac{5}{3}\Lambda_2-\frac{33}{4}\frac{\Lambda_3}{c_s^2}-\frac{8}{3}\Lambda_5-\frac{11}{3}\Lambda_6-\frac{39}{4}\frac{\Lambda_7}{c_s^4}-\frac{51}{8}\frac{\Lambda_8}{c_s^2}.
\end{align}
Now, by ignoring the terms suppressed by the slow-roll parameters, each coefficient in the cubic Galileon theory reads
\begin{align}
\Lambda_2&\simeq3\biggl(1-\frac{\sigma^2}{c_s^2}\biggr),\ \Lambda_3\simeq\frac{1}{\alpha}+\sigma-\frac{\sigma}{\alpha},
\ \Lambda_5\simeq-\frac{\alpha}{2}-\frac{3}{2}\alpha\sigma,\\
\Lambda_6&\simeq\frac{3\alpha}{4}-\frac{3}{4}\alpha\sigma,\ \Lambda_7\simeq\frac{1}{6}\frac{\sigma^2}{\alpha}-\frac{1}{6}\frac{\sigma^3}{\alpha}-\frac{1}{2}c_s^2\sigma^2,\ \Lambda_8\simeq\sigma(-1+\sigma+c_s^2\alpha),
\end{align}
where
\begin{align}
\sigma:=\frac{H\mpl^2}{\Theta},\ \alpha:=\frac{\mathcal{G}_S}{\mpl^2}=\frac{r}{16c_s^3}.
\end{align}
To obtain $f^{ss}_{{\rm NL},0}\geq\mathcal{O}(10^4)$ in the cubic Galileon theory where $c_h^2=1$, one needs $c_s\leq\mathcal{O}(10^{-6})$ (see Eq.~(\ref{eq: const_fnlnBD_fnlBD})).
Also, it can be seen from Eq.~(\ref{eq: result-fNL}) that
\begin{align}
f^{ss}_{{\rm NL},0}\lesssim 10^2\left(\frac{b_1}{\mathcal{G}_T}+\frac{b_2}{\mathcal{F}_T}\right),
\end{align}
where
\begin{align}
\frac{b_1}{\mathcal{G}_T}+\frac{b_2}{\mathcal{F}_T}=\frac{1}{4}\left(1-\sigma\right)-\frac{\sigma}{4}f_\Theta\simeq\frac{1}{4}\left(1-\sigma\right).
\end{align}
The only controllable parameter in the above is $\sigma$ in the cubic Galileon theory, and thus one may require $\sigma\geq\mathcal{O}(10^2)$ to achieve $f^{ss}_{\rm NL,0}\geq\mathcal{O}(10^4)$. For $\sigma\geq\mathcal{O}(10^2), r|_{k_{\rm CMB}}\sim 10^{-2}$, and $c_s\leq\mathcal{O}(10^{-6})$, we have $\tilde f_{\rm NL}\geq\mathcal{O}(10^{16})$, which indicates that either $f^{\rm eq}_{\rm NL}$ or $f^{\rm fold}_{\rm NL}$ (or both of them) cannot satisfy the observational bounds. In the full Horndeski theory having both $G_4$ and $G_5$, there are additional functional degrees of freedom in $\tilde f_{\rm NL}$ such as $\mu$ and $\mathcal{G}_T$, and hence the above argument is not applied to more general theories. We thus conclude that the $G_4$ and $G_5$ terms play crucial roles in enhancing the SGWB anisotropies.

\bibliographystyle{ptephy}
\bibliography{SGWB}

\end{document}